\documentstyle[aps,epsf]{revtex}

\tighten

\begin{document}

\twocolumn[\hsize\textwidth\columnwidth\hsize\csname @twocolumnfalse\endcsname

\title{Tight-binding molecular-dynamics studies of defects and disorder in
covalently-bonded materials}

\author{Laurent J. Lewis \cite{lewadd}}

\address{D{\'e}partement de physique and Groupe de recherche en physique et
technologie des couches minces (GCM), Universit{\'e} de Montr{\'e}al, Montr{\'e}al,
Qu{\'e}bec, Canada, H3C 3J7}

\author{Normand Mousseau\cite{mousadd}}

\address{Department of Physics and Astronomy, Ohio University, Athens, OH
45701, USA}

\maketitle

\begin{center}
Invited review article for {\em Computational Materials Science}
\end{center}

\begin{abstract}

Tight-binding (TB) molecular dynamics (MD) has emerged as a powerful method
for investigating the atomic-scale structure of materials --- in particular
the interplay between structural and electronic properties --- bridging the
gap between empirical methods which, while fast and efficient, lack
transferability, and {\em ab initio} approaches which, because of excessive
computational workload, suffer from limitations in size and run times. In
this short review article, we examine several recent applications of TBMD in
the area of defects in covalently-bonded semiconductors and the amorphous
phases of these materials.

\end{abstract}

\vskip2pc]
\narrowtext

\section{Introduction}

As one will be able to judge from this special issue of {\em Computational
Materials Science}, tight-binding (TB) molecular dynamics (MD) has evolved
into a powerful method for understanding material properties at the atomic
level, offering a good compromise between empirical\cite{all87} and
first-principles\cite{pay92} approaches for describing the interactions
between atoms. Indeed, empirical methods lack transferability --- model
potentials are usually fitted to specific material properties in specific
configurations, and often fail to properly describe situations and properties
other than those on which they were fitted. In contrast, first-principles
methods are transferable, but their computational workload is so great that
only small systems (less than a hundred particles or so) can be dealt with on
relatively short timescales (at most a hundred ps or so) on the fastest
computers.

With TBMD, and in particular the novel ${\cal O}(N)$ methods (for a review,
see, for instance, Ordejon's paper in this journal and Ref.\ \cite{gal96}),
it is possible to simulate systems containing several hundred particles for a
good fraction of a ns. This allows a number of interesting problems to be
addressed, as we illustrate below. In fact, the accuracy and power of the
method can be enhanced considerably by combining it with other approaches,
either empirical or first principles; we also give examples of this below.

The purpose of the present review is to illustrate the scope of application
of the TBMD method by means of selected examples. TB is a field that has been
active for some time in the world of electronic structure
calculations\cite{har89}, but only in recent years has it been coupled to MD,
making it possible to study the interplay between structure and physical
properties of materials. Thus it is possible, with TBMD, to investigate
dynamical properties {\em per se}, including relaxation, as well to optimise
structural models, after which the electronic and other properties can be
determined.

We focus here on covalently-bonded semiconductors, mostly Si and the III-V's;
carbon is the object of another article in this Journal. This review is not
meant to be exhaustive --- but rather illustrative --- and thus necessarily
incomplete; we therefore apologize to all whose work is not covered or
mentioned. Two classes of problems are examined: first, defects --- be they
localised or extended --- and, second, amorphous covalent semiconductors.
TBMD has allowed significant progress to be made in the study of defects:
because they are of quantum-mechanical nature, TB potentials are more
reliable and more accurate than empirical ones; at the same time, because
they are semi-empirical, they allow larger systems to be studied on longer
timescales than fully {\it ab initio} approaches. The same same applies to
the study of amorphous materials, where the principal difficulty is in the
proper description of the wide spectrum of highly-strained environments that
are found in these materials. Of particular interest is the relation between
structural and electronic properties which, evidently, cannot be derived from
empirical models. Before discussing these applications, we provide, for
completeness, a short overview of the methodology; more details on TBMD can
be found in, for instance, Ref.\ \cite{goringe}, as well as other articles in
this issue; an excellent discussion of MD can be found in Ref.\ \cite{all87}.


\section{TBMD}

TB is a standard method \cite{har89} for computing the electronic properties
of materials in terms of a set of parameters describing the overlap between
atomic orbitals on nearest-neighbour and, sometimes, second-nearest-neighbour
sites. In order to carry out MD or static relaxation calculations (i.e., to
compute the forces), however, it is necessary to add to the total energy a
repulsive term which includes a position-dependent electronic bonding energy
as well as ionic contributions. The total energy for TBMD simulations can
thus be written as
\[
   E_{\rm tot} = \sum_i E_i + \sum_{ij} U_{\rm rep}(|{\bf R}_i - {\bf R}_j|),
\]
where the first term on the right is the ``band-structure'' energy,
i.e., the quantum-mechanical bonding energy resulting from the overlap
of atomic orbitals, and the second term is a two-body classical
potential which accounts for all the other contributions to the total
energy. The electronic eigenvalues are obtained from a simplified
local-atomic-orbital representation of the Hamiltonian. The electronic
wave-functions are generally expanded in terms of a reduced number of
localised, orthonormal basis functions
\[
   | \psi_i \rangle = \sum_{\mu} a^i_{\mu} |\mu\rangle,
\]
where the coefficients are obtained by solving
\[
   \sum_{\nu} H_{\mu \nu} a^i_{\mu} = E_i a^i_{\mu}.
\]
In general, the basis set $|\mu\rangle$ is restricted to the valence electron
states. In the case of silicon, for example, one typically uses the single
3$s$ and the three 3$p$ orbitals --- a much smaller basis set than would be
needed in a plane-wave description of the electron wavefunctions.

Different potentials differ in the way that the Hamiltonian matrix elements
$H_{\mu\nu}$ are approximated, but also in the functional form of the
two-body potential. The matrix elements --- which depend {\it a priori} on
distance and bond angles --- are determined either from {\it ab initio}
calculations or extracted from experiment. The most common approximation
consists in parametrising the overlap integral in terms of a set of constants
which decay as $1/r^2$, with a cutoff distance between the first and
second-neighbour shells; a short cutoff distance speeds up the calculations
but can be a source of problems in disordered materials where near-neighbour
shells are not perfectly separated. It is also possible to compute the matrix
elements and the repulsive potential in a more accurate way using either the
local-density approximation \cite{porezag95} or the local-orbital
representation of density-functional theory first introduced by
Harris\cite{harris}. These schemes, often dubbed ``{\it ab initio} TB'',
constitute a trade-off between the accrued computational effort associated
with them and the more accurate description of strained environments which is
of particular importance for disordered materials and defects.


\section{Defects}
\label{sec:defects}

Defects play a major role in determining the physical properties of
semiconductors\cite{lan81}. Even when present at low density, they affect
deeply the electronic structure of the materials. This is true of point
defects, but also of extended defects, especially in the case of quantum
structures. In spite of numerous experimental or theoretical studies, a
complete picture of the structure of even the simplest defects (vacancies,
interstitials, and small complexes of them, such as divacancies), in the
most-studied semiconductor material --- silicon --- has not yet emerged.
Defects, further, are not static objects, and can undergo diffusion at
sufficiently high temperature. Again, little is known of such processes; the
diffusion coefficient of H in Si, for instance, is still not understood in
detail.

TBMD calculations of defects have contributed to our understanding of their
static and dynamic properties, but they have also been extremely useful in
validating the models. Indeed, because they break the local symmetry, which
often results in subtle relaxation and electronic effects, defects are
difficult to treat using empirical models and therefore serve as an excellent
test of the ability of TB models to deal with low-symmetry situations. The
details of the atomic structure of defects, however, are generally not known
from experiment, and the tests must be against {\em ab initio} calculations,
which themselves carry significant uncertainties because of limitations in
size and computational load.

Here we examine recents results on intrinsic (or native) defects. Extrinsic
defects, i.e., impurities, are a difficult problem because of the additional
complexity involved in constructing multi-species interactions. We
nevertheless consider some exceptions, notably H diffusion in Si and B
relaxation in Si.


\subsection{Intrinsic point defects in silicon}

\subsubsection{Basics}

One of the first applications of TBMD was the study, by Wang {\it et
al.}\cite{wan91}, of native point defects in crystalline silicon using the TB
model of Goodwin, Skinner and Pettifor (GSP)\cite{goo89}. Wang {\it et al.}\
have examined the formation energies of neutral monovacancies and
self-interstitials as a function of the size of the simulation cell --- up to
512 atoms. The results are listed in Table \ref{sidfe}: One sees that size
somewhat influences the formation energy, especially for the monovacancy,
indicating that the distortion pattern of the defects cannot be accommodated
fully by a small cell. The relaxed values of the formation energies seem to
fall close to results obtained in the local-density approximation (LDA)
\cite{bar84,kel92,seo96}, probably within the uncertainties inherent to both
approaches.

For the single vacancy, Wang {\it et al.}\ find a tetragonal Jahn-Teller
distortion on top of the radial displacement of nearest-neighbour atoms, as
was also observed by Song {\it et al.}\cite{son93}. This agrees with the
electron paramagnetic resonance measurements of Watkins\cite{wat86}, as well
as the LDA calculations of Baraff {\it et al.}\cite{bar80} and, more
recently, Seong and Lewis\cite{seo96}. The LDA calculations, however, predict
slightly larger total displacements than TB --- 0.4 versus about 0.3 \AA. The
Jahn-Teller distortion is of quantum-mechanical origin and therefore not
available from empirical models.

Song {\em et al.}\cite{son93} have used the GSP model to investigate, in
addition, the simple and split divacancies, as well as the Frenkel pair, all
in their neutral charge states; the results as shown in Table \ref{sidfe}.
The formation energies are large and the relaxation energies --- the
difference in energy between unrelaxed and relaxed configurations, given in
parenthesis in Table \ref{sidfe} --- clearly non-negligible, typically
representing a good fraction (30\% or so) of the formation energy. Of course,
this is accompanied by a significant change in volume during relaxation, and
atomic displacements that can be as large as 1.25 \AA\ (in the case of the
split divacancy\cite{son93}).

Within the GSP-TB model it is energetically favourable for two vacancies to
``coalesce'', saving about 1.68 eV in the process. Indeed, the formation
energy of two isolated vacancies is 7.36 eV, dropping to 6.54 eV for the
split divacancy, and to 5.68 for the simple divacancy. Thus, divacancies are
expected to readily form and be relatively stable even at high temperatures.
Likewise, the formation of a Frenkel pair by a vacancy and an interstitial
can reduce their total energy by as much as 1.43 eV -- from 7.98 to 5.55 eV.

The activation energy for diffusion is the sum of formation and migration
energies. The latter is the energy at the transition state between two
equilibrium sites. Song {\em et al.}\cite{son93} estimate the migration
energy for the vacancy to be less than 1.0 eV so the activation energy, using
the formation energy values discussed previously, must be less than 4.7 eV.
Likewise, tetrahedral interstitials migrate via hexagonal sites with an
energy of about 0.63 eV, and thus the activation energy in this case would be
of the order of 5.0 eV.

As mentioned above, the atomic displacements for the monovacancy are
predicted by the LDA to be slightly larger than the TB values. The opposite
is true in the case of divacancies, where the LDA predicts displacements
substantially smaller than the TB model of GSP\cite{seo96}. The LDA, further,
leads to a resonant-bond Jahn-Teller distortion (as opposed to the usual
pairing configuration) for the simple divacancy that is not observed in TB
calculations. Also, the relaxation energies obtained from the GSP-TB model
for the divacancies are significantly larger than the corresponding LDA
values.

The formation volumes of the vacancy and the interstitial have been
calculated by Tang {\em et al.}\cite{tan97} using the TB model of Kwon et
al.\cite{kwo94}. The formation volume is defined as $\Delta\Omega=V_{\rm rel}
\pm \Omega$, where $V_{\rm rel}$ is the relaxation volume associated with the
defect (i.e., arising from the relaxation of the atoms in the neighbourhood
of the defect) and $\Omega$ is the volume per atom of the perfect crystal;
the plus sign is for vacancies while the minus sign applies to interstitials.
Using a 216-atom supercell, and after a careful search for the equilibrium
volume of the perfect crystal, Tang {\em et al.}\ obtained a relative
formation volume $\Delta\Omega/\Omega$ of 3\% (contraction) for the vacancy
and $-$10\% (expansion) for the interstitial. Thus, the volume changes
arising from the presence of these defects should cancel each other to a
large extent, in agreement with diffuse x-ray scattering
experiments\cite{bau99}.

Though the picture is far from being complete, and it is therefore difficult
to draw meaningful conclusions on the accuracy of the TB models, it seems to
be the case that the model of Kwon {\em et al.}\cite{kwo94} overestimates the
relaxation energies, while that of Lenosky {\em et al.}\cite{len97} appears
to be doing better (at least for the monovacancy). Bernstein and
Kaxiras\cite{ber97} have observed that the agreement between TB and LDA
defect formation energies can be improved significantly by relaxing the
constraint on the band gap, which is then allowed to vary in the fitting
process. Clearly, more precise TB models are necessary in order to capture
the subtle details of such low-symmetry situations. Also, more (and
better-converged) first-principles calculations are needed to provide a
proper reference database for comparison.

Rasband {\em et al.}\cite{ras96a} have studied the convergence of intrinsic
defect formation energies with respect to potential cutoff distance as well
as number of points used to sample the Brillouin zone. Considering isolated
vacancies as a test case, they found these variables to affect only very
slightly the {\em unrelaxed} formation energy, while the effect of relaxation
can be sizable. For instance, increasing the cutoff from 3.2 \AA\ (between
first and second neighbours) to 4.1 \AA\ (between third and fourth) and using
40 {\bf k} points rather than one causes the vacancy formation energy to
decrease from 3.67 (cf.\ Table \ref{sidfe}) to 3.15 eV. The corresponding
values for the $-$, $+$ and $2+$ charge states of the vacancy are 2.9, 3.6
and 4.1 eV, respectively. The $+$ vacancy is nowhere in the gap a favourable
state of the defect, and thus gives rise to the so-called ``negative-$U$''
effect, that is an effective correlation energy between electrons which is
negative\cite{and75} (see also below for GaAs).

For the tetrahedral interstitial in its neutral state, now, Rasband et
al.\cite{ras96a,ras96b} find a formation energy of 4.7 eV, using a 4.1 \AA\
cutoff and 40 {\bf k} points. This is comparable to the 4.40 eV value given
in Table \ref{sidfe} (3.2 \AA\ cutoff, $\Gamma$-point only). In the $-$, $+$,
$2+$, and $3+$ charge states, the corresponding numbers are 5.5, 4.2, 3.5,
and 4.1 eV, respectively, taking the Fermi level in the middle of the gap.
Thus, $2+$ interstitials should occur with a much larger probability than
other charge states. This prediction of the stability of the $2+$
interstitial is in agreement with earlier {\em ab initio} results and might
explain the discrepancy between experiment (or more precisely
``model-fitted'' experimental data --- cf.\ Fig.\ 2 in Ref.\ \cite{ras96b})
and many calculations. In particular, this result is consistent with metal
in-diffusion experiments (see \cite{ras96b} for references).

Rasband {\em et al.}\cite{ras96b} have used TBMD to search for new defect
structures (interstitials) and have found a whole family of them. For the
neutral single interstitial, three stable configurations are found, viz.\ the
T interstitial (formation energy of 4.7 eV), the 110-split interstitial (5.0
eV) and the 100-split interstitial (5.4 eV). The corresponding ionised
defects with charges between $+3$ and $-2$ range in energy from 5.0 to 6.3 eV
for the 110 split, and from 4.5 to 6.3 for the 100 split. As we have seen
above, the doubly-ionised T interstitial has a formation energy of 3.5 eV and
is thus much more stable than any of the split interstitials.
Di-interstitials were also examined. Rasband {\em et al.}\cite{ras96b} found
a novel low-energy configuration --- the ``split triple'' interstitial,
consisting of three Si atoms sharing one lattice site and forming an
equilateral triangle in a (111) plane. The formation energy of this defect is
a small 3.65 eV per atom in the neutral state, dropping to 3.0 eV in the $2+$
state. A similar 110 split-triple interstitial has a 3.3 eV formation energy,
while a ``Z'' configuration has 3.4 eV, both in their $2+$ state, which is
the most stable. There is, to our knowledge, no evidence that these objects
have been observed experimentally, nor are there other calculations to
compare with. In view of the approximate character of TB, the stability of
such defects should probably be considered as somewhat speculative at this
point.

\subsubsection{Energy levels}

One advantage of TB over empirical approaches is that it gives access to the
electronic structure of the material. The GSP model does not provide a very
good description of the band structure of Si. It is nevertheless informative
to examine, within this model, the effect on the electronic properties of the
presence of defects. Fig.\ \ref{sidefdos} shows the density of states near
the band gap for the various defect types considered by Song et
al.\cite{son93}. Defects, evidently give rise to localised bands in the gap.
Table \ref{sideflev} lists the positions of the levels associated with the
defects in their relaxed configurations.

The LDA results quoted in Table \ref{sideflev} are those of Ref.\
\cite{seo96}, where other references can also be found; however, only very
few calculations of the defect levels have been reported. For the
monovacancy, for instance, the LDA value for the highest occupied level, 0.23
eV, is in rough agreement with the self-consistent-field calculations of
Lipari {\em et al.}\cite{lip79}, which gives a singlet level at 0.3 eV (but
relaxation was not taken fully into account). Both calculations however
disagree with the TB result. For the divacancies, the error bar on the LDA
calculation (about 0.1 eV) is such that it is not clear that this defect
leads to levels in the gap, in apparent disagreement with the TB results. It
should be said, as discussed in Ref.\ \cite{seo96}, that the precise
positions of defect-induced gap levels are rather sensitive to the details of
the computational model. All calculations, however, point to the fact that
defect levels move towards the valence-band maximum upon relaxing the
structure.

Electron paramagnetic resonance (EPR) experiments gives information about the
energy levels of charged defects. The $+$ and $2+$ charged states of the
vacancy lie close to the top of the valence band, at 0.05 and 0.13 eV,
respectively, whereas the $-$ and $2-$ lie at 0.57 eV and 0.11 eV below the
bottom of the conduction band. The TBMD results of Rasband et al. are in
error with experiment by $-$0.04, $-$0.07, $-$0.17, and $-$0.90 eV,
respectively (see \cite{ras96a} for experimental references). Except for the
$2-$ defect, the agreement can be considered as good.

\subsubsection{Vacancy clusters in silicon}

Structural evolution during non-equilibrium processes such as irradiation and
growth is mediated, to a large extent, by the presence of defects or clusters
of them. Vacancies, for instance, may coalesce and give rise to microdefects
such as voids, bubbles, or dislocations. It is therefore important to have
some idea of the structure and energetics of such imperfections.

The case of vacancy clusters has been examined by Bongiorno, Colombo, and
Diaz de la Rubia\cite{bon98} (BCDR) using TBMD. This work was motivated to a
large extent by serious disagreements between model potential
(Stillinger-Weber) and first-principles studies with regards, in particular,
to the shape and energetics of stable clusters. In order to perform these
calculations, BCDR exploited the power of the Goedecker-Colombo linear
scaling scheme\cite{goe94} as implemented on a (parallel) Cray T3E. This
allowed very large simulation cells to be considered --- 1000 atoms ---
containing vacancy clusters varying in size between 1 and 35. In all cases,
the lowest-energy configurations was obtained through a careful relaxation.

BCDR examined two different cluster shapes: spheroidal clusters (SPC), where
vacancies are created by simply removing atoms from successive radial shells
about a single vacancy, and ``hexagonal ring clusters'' (HRC), where Si atoms
are removed following a ring pattern in the crystal. These structures
evidently are different from the geometric viewpoint, but also from the
``chemical'' viewpoint. Indeed, for a given cluster size, SPC structures have
more dangling bonds than HRC structures, as demonstrated in the top panel of
Fig.\ \ref{sivaclus}, and SPC are therefore expected to be less favourable
than HRC because dangling bonds are expensive. This is in fact confirmed by
the formation energy, middle panel in Fig.\ \ref{sivaclus}, at least for
clusters containing 1--24 atoms. For larger clusters, the aggregation path is
likely very complex and probably depends on the details of the kinetics of
the non-equilibrium process (irradiation or growth).

The binding energy for the various clusters is given in the bottom panel of
Fig.\ \ref{sivaclus}; there are clearly some ``magic clusters'', i.e.,
clusters who are much more stable than others slightly different in size.
This is the case, for instance, of HRC clusters with $n= 6$, 8, 10, 12, 14,
16, 18, etc.; these magic clusters are the result of minimising the number of
dangling bonds as as well as structural rearrangements (internal
reconstructions).


\subsection{Impurity levels in Si and GaP}

As we have seen, relaxation affects strongly the structure and energetics of
intrinsic defects in semiconductors. This is true also of impurities. The
deep levels associated with several impurities in Si and GaP have been
investigated using TBMD by Li and Myles\cite{lim91}. Their approach is based
on the theory of Hjalmarson {\em et al.}\cite{hja80} of deep levels,
augmented to include lattice relaxation by adding a repulsive pair potential
derived from Harrison's overlap interactions\cite{har83}. The host material
is described using the $sp^3s^\star$ TB model of Vogl {\em et
al.}\cite{vog83}. The relaxation is performed {\em via} MD, but restricted to
nearest neighbours and $T_d$ symmetry-conserving displacements, i.e.,
breathing modes. Full details of the method can be found in Ref.\
\cite{lim91}.

Table \ref{sigap} gives the energies of several deep-level impurities in Si
and GaP as calculated by Li and Myles\cite{lim91}. These are of $A_1$ (as
well as $T_d$) symmetry, i.e., $s$-like. Also given in Table \ref{sigap} are
values from experiment (see \cite{lim91} for references) and the results from
the Hjalmarson {\em et al.}'s theory\cite{hja80}, which does not include
relaxation. Clearly, relaxation is sizable in all cases and improves quite
significantly the agreement with experiment. Relaxation proceeds inwards in
all cases except GaP:O. Li and Myles indicate that inclusion of
second-neighbour relaxation changes the results very little.


\subsection{Boron in silicon}

Boron is a dopant which is routinely used in the semiconductor fabrication
process and it is therefore important to understand at the fundamental level
how the host material is affected by the impurity, and how the latter
diffuses. As a first step in this direction, Rasband et
al.\cite{ras96a,ras96b} have used TBMD to study defect-dopant pairs as well
as boron interstitials in silicon. For Si-Si interactions, the GSP model was
employed. For Si-B interactions, a new model {\em \`a la} GSP was developed,
with the parameters determined by fitting to {\em ab initio} band-structure
energies for zinc-blende SiB; full details can be found in Ref.\
\cite{ras96a}.

The TBMD results are found to be generally in good agreement with the {\em ab
initio} calculations of Nichols {\em et al.}\cite{nic89}. For the boron T
interstitial, the TB model gives a formation energy of 3.7 eV, compared to
3.9 eV from first principles. For the neutral boron-substitutional--vacancy
complex, Rasband {\em et al.}\ find 2.9 eV compared to 3.0 for Nichols {\em
et al.}; the binding energy of this complex is $-$0.39 eV, which compares
well to $-$0.5 {\em ab initio}. The binding energy of the
boron-substitutional--interstitial complex is $-$1.2 eV from TBMD, in line
with model-fitted data, $-$1.5 eV. Relaxation of the four Si atoms near a B
substitutional impurity is found to be small --- about 0.04 \AA, independent
of the state of charge.

Rasband {\em et al.}\cite{ras96b} have looked for ``new'' defects involving
boron. Starting with a boron atom in a T-interstitial position with charge
states in the range $-2$ to $+2$, they identified a series of seven different
interstitial species having formation energies, in their most stable charge
states, between 3.1 and 5.5 eV. They also examined the case where the
starting structure was a combination of Si 110-split interstitial adjacent to
a B T-interstitial; the resulting configuration after relaxation was a
110-split di-interstitial with a formation energy of 2.5 eV per atom.
Further, they observed the formation of a ``ring'' di-interstitial having a
formation energy of 3.1 eV per atom exhibiting a distinctive five-membered
ring.

The above results, just as was the case for native defects in silicon,
indicate that new elements must be taken into account when experimental
defect concentration data are analysed. In particular, assumptions regarding
the prevailing charge state of any particular defect should probably be
re-examined at the light of the above (and other) results.


\subsection{Point defects in GaAs}
\label{sec:gaas}

In the above study of impurity levels in Si and GaP, Section B, relaxation
was restricted to radial displacements. While such distortions might
constitute the most important component of relaxation, it may be expected
that, quite generally, the $T_d$ symmetry is broken by the presence of
defects, and this can only be revealed through full relaxation of the
lattice.

A TBMD investigation of defect relaxation and energetics in GaAs has been
proposed by Seong and Lewis\cite{seo95}. The study is based on the model of
Molteni, Colombo, and Miglio\cite{mol93}, which takes charge-transfer effects
into account through a screened, and thus short-range, Coulomb term. Native
point defects in GaAs play an important role; this is true, in particular, of
the so-called ``EL2'' defect, which can compensate residual acceptor
impurities and pin the Fermi level at midgap, so that GaAs can be
manufactured without intentional impurity doping. This defect is thought to
be related to the As antisite (substitution of a Ga by an As).

For tetragonally-distorted configurations, it is convenient to describe the
relaxation in terms of three displacement components\cite{lan81,laa92}: the
``breathing'' (radial) mode, and two ``pairing'' modes, which measure the
lateral displacements of the ions (i.e., perpendicular to the breathing
mode). Table \ref{rlxb} gives the breathing-mode displacements for Ga and As
vacancies and antisites in various states of charge. The pairing modes are
negligibly small for Ga vacancies and As antisites, but are sizable for As
vacancies and Ga antisites; they are given in Table \ref{rlx}.

Ga vacancies are found to relax inwards by an amount which is independent of
the state of charge, leading to an open-volume contraction of about 34\%,
consistent with positron lifetime measurements{\cite{saar91,saar}, as well as
first-principles calculations\cite{laa92}, though relaxation in the latter
case is substantially smaller and the position of the corresponding localised
states different. The relaxation of As antisites also is the same for all
charge states; it is small and, in contrast to Ga vacancies, proceeds
outwards, leading to a volume expansion of about 7.5\%. This is also what is
found in first-principles calculations\cite{dabr,chadi,bach} and
experiment\cite{feens}. We find the localised state for this defect to lie
0.49 eV above the valence band maximum, in accord with tunneling spectroscopy
measurements\cite{feens}.

Both As vacancies and Ga antisites undergo tetrahedral-symmetry-breaking
distortions, i.e., have non-zero pairing displacements. In both cases,
further, the relaxation pattern depends sensitively on the state of charge,
as can be seen in Tables \ref{rlxb} and \ref{rlx}. For As vacancies, changes
in the open volume range from a 54\% expansion in the $+$ state to a small
3\% contraction in the $2-$ state. The positron lifetime is therefore
expected to be longest in the $+$ state and shortest in the $2-$ state; this
is consistent with experiment, which finds a decrease from 295 to 258 ps when
the charge changes from neutral to singly negative\cite{saar91}. In the case
of Ga antisites, the changes in the open volume vary between +50\% and a
negligible $-$1\%; the latter is for the doubly charged vacancy and
corresponds, approximately, to the GaAs crystal in its ``normal'' state. (An
As with five valence e$^-$ is replaced by a Ga with three e$^-$, then two
electrons are added to it).

It is of interest to examine more closely the structure of the Ga antisite.
The singly-negative Ga antisite undergoes a small ($<$2\%) outward
breathing-mode displacement, and similar pairing-mode relaxations of the
nearest neighbours. The antisite itself is displaced slightly from its ideal
lattice position [by about 0.1 \AA\ in the $(11\bar 1)$ direction]. The
distance between the antisite and one of its Ga neighbours (referred to as
``Ga$_1$'' by Zhang and Chadi\cite{zha90}) is 2.65 \AA, a little bit longer
than the equilibrium bond length (2.45 \AA). The distance between the Ga
antisite and the three other Ga neighbours, which relax very little, is 2.44
\AA. This atomic structure near Ga$_{\rm As}^{-}$ is very similar to that of
Zhang and Chadi\cite{zha90} who studied the problem from first principles;
the configuration is depicted in Fig.\ \ref{ga_antisite}(a). The relaxed
state of the {\em neutral} Ga antisite is shown in Fig.\
\ref{ga_antisite}(b). The defect moves away by 0.4 \AA\ from its equilibrium
position; the Ga$_1$ atom also moves away, but in the opposite direction, by
0.48 \AA. This leads to a ``broken-bond'' configuration with a relaxed
Ga$_{\rm As}^0$--Ga$_1$ distance of 3.32 \AA, in perfect agreement with Zhang
and Chadi's calculation\cite{zha90}. Positively-charged Ga antisites show
similar relaxations.

The variation with electron chemical potential ($\mu_e$, where $0 \leq \mu_e
\leq E_g$ and $E_g$ is the width of the gap) of the formation energy of a
given defect in various states of charge provides information about its
ionisation level and most stable state. This is displayed in Fig.\
\ref{gaasdfe} for the four types of defects considered. A detailed discussion
of these results can be found in \cite{seo95}. The formation energy also
depends, however, on the atomic environment that prevails during growth,
i.e., on atomic chemical potentials. Fig.\ \ref{df_mga} shows, for three
different values of $\mu_e$, the defects of lowest energy as a function of
the deviation $\Delta\mu_{\rm Ga}$ from the bulk value. At both the
valence-band maximum (VBM) and midgap, the As antisite is the most favourable
defect in the As-rich limit, while the Ga antisite is the one preferred in
the Ga-rich limit. In the middle of the $\Delta\mu_{\rm Ga}$ range, the
vacancies $V_{\rm As}^+$ and $V_{\rm Ga}^{3-}$ take over at midgap, though,
in only a narrow window of values. At the conduction-band maximum (CBM), now,
$V_{\rm Ga}^{3-}$ dominates in As-rich conditions, while Ga$_{\rm As}^{2-}$
is favoured in the Ga-rich limit. Thus, overall and by in large, antisites
have the lowest energies, and these defects are expected to be much more
prevalent than vacancies in GaAs. This is consistent with the current
interpretation\cite{dabr,cha88,bar85,bar86,saa94,sie93} of the EL2 defect
being a neutral As antisite with tetrahedral symmetry.


\subsection{Dynamics and kinetics of point defects}

\subsubsection{Vacancy and interstitial diffusion and recombination}

Vacancies and interstials (as well as small complexes of them and impurities)
contribute strongly to the transport of mass in crystalline solids. Yet,
little is known on their equilibrium concentrations and diffusivities. The
values from experiment vary widely, and it is not clear which of vacancies or
interstials dominate transport at a given temperature. Further, the
recombination process of the two species is not well understood, and in
particular it is not clear whether or not there exists an energy barrier
against the recombination. These questions have been addressed, in part,
using MD and empirical potentials (e.g., Stillinger-Weber
\cite{stillinger85}); such calculations however cannot account for the subtle
quantum-mechanical details of the process. First-principles methods, on the
other hand, are hampered by limitations in size and time that make the study
of such problems almost impossible at this time. TBMD offers a good
compromise between the two, and was used by Tang {\em et al.}\ to study
vacancy and interstitial diffusion and recombination\cite{tan97}. For this
problem, they employed the TBMD model of Kwon {\em et al.}\cite{kwo94}, which
appears to be more accurate than the GSP model insofar as the properties of
point defects are concerned for a 216-atom model. The diffusivity
calculations were however performed on a 64-atom (plus or minus one) system,
running for as long as 200 ps.

The diffusion constant is given by the product of diffusivity and equilibrium
concentration, $D(T)=d(T)C(T)$, where
\[
   d(T)=d_0\exp(-E_m/k_BT)
\]
and
\[
   C(T)=\exp[-(E_f-TS_f)/k_BT].
\]
$E_m$ is the migration energy and $E_f$ is the formation energy. If the
entropy of formation, $S_f$, is assumed to be independent of temperature,
then the diffusion constant takes the Arrhenius form:
\[
   D(T)=D_0\exp(-E_a/k_BT)
\]
where $D_0=d_0\exp(S_f/k_B)$ and $E_a=E_m+E_f$. $D(T)$ is indeed observed in
crystalline silicon to be Arrhenius over a wide range of temperatures. The
precise values of the prefactor $D_0$ and activation energy $E_a$ for the
individual species are still a matter of discussion, but the following values
have recently been given by G\"osele {\em et al.}\cite{gos96}: $E_a^V=4.03$ eV
(vacancies), $E_a^I=4.84$ eV (interstitials), and
$D_0^I/D_0^V=1.53\times10^4$.

The diffusivities of vacancies and interstitials were simulated explicitly by
Tang {\em et al.}\cite{tan97} using TBMD. The migration energies were found
to be $E_m^V=0.1$ eV and $E_m^I=1.37$ eV. In this model, the formation
energies are $E_f^V=3.97$ eV and $E_f^I=3.80$ eV, leading to activation
barriers of $E_a^V=4.07$ and $E_a^I=5.17$ eV, in remarkable agreement with
the experimental values quoted above. Tang {\em et al.}\ did not calculate
the entropies of formation --- this is a very difficult calculation --- but
have fitted them to first-principles calculations and experiment; they obtain
$S_f^V=9k_B$ and $S_f^I=11.2k_B$. The resulting prefactors are $D_0^V=0.96$
cm$^2$/s and $D_0^I=1.16\times10^4$ cm$^2$/s, i.e.,
$D_0^I/D_0^V=1.21\times10^4$, again in agreement with experiment. Arrhenius
plots of the diffusion constants are given in Fig.\ \ref{ivdiff}. Diffusion
is dominated by vacancies at low temperature. At high temperature, above 1080
$^\circ$C, in spite of a sizeably larger migration barrier, interstitials
take over. This crossover effect is due to the relative values of the
prefactors --- $D_0^I$ is four orders of magnitude larger than $D_0^V$ ---
and find it's origin in the Meyer-Neldel (compensation) law, which states
that, for a family of activated processes, the prefactor increases
exponentially with the activation barrier. The validity (and origin) of the
Meyer-Neldel law was established in the case of surface diffusion by Boisvert
{\em et al.}\cite{boi95}.

Mass transport, as noted above, will evidently be affected by the rate at
which vacancies and interstitials recombine, which itself depends on the
existence or not of a recombination barrier. Tang {\em et al.}\cite{tan97}
have used TBMD to investigate this problem. Starting with an
interstitial-vacancy pair with the two defects separated by some distance ---
between two and six nearest-neighbour spacings --- MD simulations were
carried out at finite temperature in order to follow the migration path and
eventual recombination of the defects. The latter does not always take place.
For instance, using as initial configuration a $<110>$ dumbbell interstitial
and a vacancy three nearest-neighbour spacings away, Tang {\em et al.}\ found
that recombination did not occur at 300 K: thermal activity at this
temperature is not large enough to overcome the local distortions around the
dumbbell induced by the close approach of the vacancy, at least on the
timescale of the simulations. Instead, an ``I-V complex'' forms; the
quenched-in state of the complex, which has a formation energy of 3.51 eV, is
shown in Fig.\ \ref{ivcomp}(a). At elevated temperature, however, the I-V
complex annihilates. This is shown in Fig.\ \ref{ivcomp}(b)-(e). The
annihilation path is found to be a bond-switching process and takes place in
about 10 ps at 1500 K; the activation (migration) barrier $E_b$ for the
process is about 1.23 eV. The lifetime $\tau$ of the complex is roughly given
by $\tau^{-1}=\nu\exp(-E_b/k_BT)$; taking $\nu$ as the Debye frequency,
$\sim10^{13}$ Hz, one finds $\tau$ to be of the order of hours at room
temperature, but only a few microseconds at typical annealing temperatures
(i.e., 700-800 K or so). The recombination of vacancy-interstitial pairs, in
particular with regards to the electronic structure, was examined by Cargnoni
{\em et al.}\cite{car97,car98} by a combination of TBMD and {\em ab initio}
Hartree-Fock calculations.

\subsubsection{Hydrogen diffusion in silicon}
\label{sec:hydrogen}

Because it may form complexes with a variety of intrinsic or extrinsic
defects, hydrogen in semiconductors affect deeply their optical and
electronic properties. It is usually present as a result of the complicated
fabrication process, but is also often intentionally included in order to
passivate defects (e.g., unsaturated, or ``dangling'', bonds). Being a light
species, hydrogen diffuses readily, inducing additional defects along the way
and thus affecting the transport properties of the material to an extent
which is determined by its relative concentration. It is therefore important
to understand diffusion at the atomic level in order to gain better control
on the properties of semiconductors.

The rate of diffusion of hydrogen in crystalline silicon remains a matter of
debate in spite of the simplicity of the structure of the host material. At
low temperatures (below 800 K), the experimental values that have been
reported in the literature vary widely, sometimes by as much as two orders of
magnitude at a given temperature (see, e.g., \cite{pan94} for references) .
There are very few data available at high temperatures (above 1000 K), but
they are in reasonable agreement with the {\em ab initio} MD simulations of
Buda {\em et al.}\cite{bud89} for H$^+$ (proton) diffusion. Unfortunately,
because the timescale for diffusion increases exponentially with temperature,
{\em ab initio} simulations cannot be carried out at lower temperatures and
there exists no empirical model that can deal with this system in a
sufficiently accurate way.

The problem of hydrogen diffusion in silicon was addressed using TBMD by
Panzarini and Colombo (PC)\cite{pan94} as well as by Boucher and DeLeo
(BDL)\cite{bou94}. This constitutes an interesting application of the method
as it extends quite significantly the range of temperatures that were covered
by the {\em ab initio} simulations of Buda {\em et al.}\cite{bud89} while
accounting, still, for the quantum-mechanical nature of the system. For
Si--Si interactions, both used the GSP model; for other interactions, the
appropriate parameters were determined by fitting to selected properties of
the silane (SiH$_4$) molecule.

Both models give the bond-center (BC) site as the equilibrium site in static
conditions (i.e., at 0 K), in agreement with electron paramagnetic resonance
\cite{gorxx} and muon spin resonance experiments \cite{kie86}, as well as
previous first-principles calculations (see, e.g., \cite{bud89}). In the BDL
model, this accord is obtained at the expense of a phenomenological parameter
that accounts for differences in environment between silane and crystalline
Si. Other equilibrium sites, almost degenerate in energy (differences of less
than $\sim$0.1 eV) with the BC site, are also found. PC, for instance, find
the hexagonal interstitial site to be very close in energy to the BC site.
Both calculations predict that when H sits in the BC position, nearby Si
atoms relax by about 0.4 \AA, while very little relaxation is observed when H
is in the hexagonal site; this agrees with first-principles calculations.

The diffusion path is also subject to debate. Though experiment and theory
agree that the lowest energy site is the BC site, the energy of nearby
metastable sites --- which likely play an important role in diffusion --- is
not known precisely. Details of the diffusion path, further, are expected to
be strongly affected by dynamical effects because the heavy Si atoms cannot
follow adiabatically the motion of the lighter H atom. The {\em ab initio}
simulations of Buda {\em et al.}\cite{bud89}, which cover a timescale of
about 4 ps, suggest that diffusion consists of a series of hops between
highly-symmetric interstitial sites, but other possibilities can also be
envisaged.

Both PC and BDL carried out their TBMD simulations for a single H in a
64-atom {\em c}-Si supercell. The simulations by BDL cover the temperature
range 1050--2000 K, and run for a maximum of 42 ps, while PC examined
temperatures in the range 800--1800 K, running their simulations for as long
as 300 ps at the lowest temperatures.

Fig.\ \ref{hsidiff} presents the results of several measurements of the
diffusion constant, plotted in the manner of Arrhenius. Also indicated on
this plot are the {\em ab initio} MD data points of Buda {\em et
al.}\cite{bud89}; they are found to agree (at least qualitatively) with the
high-temperature experimental points of Van Wieringen and Warmholtz, which
can be fitted to an Arrhenius law, $D(T)=D_0\exp(-E_A/k_BT)$, with
$D_0=9.41\times10^{-3}$ cm$^2$/s and $E_A=0.48$ eV. When extended to low
temperatures (dashed line in Fig.\ \ref{hsidiff}), one clearly sees the
deviations from the Arrhenius behaviour; it should be said, however, that
there is no ``guarantee'' that diffusion should be Arrhenius over the whole
range of temperatures.

The TBMD data of PC are also indicated in Fig.\ \ref{hsidiff}. The agreement
with experiment is clearly excellent in the high temperature limit. The data
of BDL are not plotted in this figure, but they are found to be extremely
well fitted by the Arrhenius law with $D_0=6.91\times10^{-3}$ cm$^2$/s and
$E_A=0.45$ eV all the way down to 1050 K. This is in striking agreement with
experiment, as can be judged by the close similarity between the prefactors
and energy barriers. (The differences are unsignificant). PC however observe
deviations from Arrhenius already at 1200 K. The reason for this minor
discrepancy with the BDL data can perhaps be traced down to statistics.
Indeed, 300 ps remains relatively short on diffusion timescales. For
instance, for a diffusion constant of $10^{-6}$ cm$^2$/s, a quick calculation
indicates that the average distance visited by a diffusing particle would be
about 4 \AA. This might explain, in part, the deviations that are seen at
even lower temperatures; clearly, however, the TBMD data are consistent with
the low-temperature behaviour seen in experiment. Further calculations are
evidently necessary to reconcile theory and experiment at low temperatures.

Detailed analysis of the trajectories of individual atoms makes it possible
to elucidate the mechanism for diffusion. Fig.\ \ref{hsipath} shows the
diffusion path of the H atom over a 25-ps period at 1200 K as calculated by
PC; BDL obtain very similar results. The H atom diffuses preferentially via a
sequence of jumps from one BC site to another, spending very little time in
between; this corresponds to a low-memory, high-friction regime, where the
directions of consecutive jumps are uncorrelated. It is found, further, that
the H atoms avoids the low charge-density regions, which is inconsistent with
the observations of Buda {\em et al.}\cite{bud89} for the diffusion of H$^+$.
Both PC and BDL find a correlation between diffusion events (jumps) and the
vibrations of nearby Si atoms: At low temperature (less than 850 K or so),
the Si--Si stretching mode ($\sim$65 meV) is not thermally excited, and
diffusion becomes difficult because the BC site is an equilibrium site for
hydrogen. At these temperatures, further, long jumps (i.e., to sites more
distant than near-neighbours) are quenched in.


\section{Disordered phases}
\label{sec:amorphous}

As already noted in the Introduction, disordered materials ---
covalently-bonded semiconductors and chalcogenides --- is another area where
TBMD simulations have provided important new insights, in particular with
regards to the interplay between structure and electronic properties.

Many empirical potentials have been developed that give a reasonable
description of the liquid and crystalline phases of semiconductors, as well
as, for some, clusters; this is the case for instance of the models of
Stillinger and Weber\cite{stillinger85}, Tersoff\cite{tersoff88}, Biswas and
Hamann\cite{biswas87} and, recently proposed, Bazant {\it et
al.}\cite{bazant97} These models however lack the generality to account for
the large variety of highly strained environments --- caused by elastic,
topological or chemical disorder --- that are encountered in amorphous
materials. Thus, for instance, the Stillinger-Weber potential fails to
reproduce the clean separation observed experimentally between first and
second neighbour peaks in the radial distribution function. The problem is
particularly serious for materials such as amorphous carbon, where atoms can
be found in several different bonding states, or compound materials, where
bonding is partly ionic and the nature of the chemical environment plays an
important role. There exists, for example, no satisfactory empirical
potential for the III-V amorphous compounds.

The availability of accurate potentials is a necessary, but not sufficient,
condition for constructing structurally sound amorphous models. The
``preparation'' algorithm must be capable of finding a reasonable ground
state --- one in which the number of topological and electronic defects is a
minimum --- i.e., relaxation must be as exhaustive as possible.
Unfortunately, this is more difficult with TB than empirical potentials.

The strong intermediate range disorder found in amorphous semiconductors
poses an additional challenge for the development of semi-empirical TB
potentials. As for the liquid phase, near-neighbour shells are often
ill-defined: The first and second neighbour shells can be blurred either
through disorder or simply large mismatch, as is the case for InP, for
example. Attention must therefore be given to the radial behaviour of the
empirical repulsive potential as well as the cutoff of the matrix element
interactions. In spite of these concerns, current semi-empirical potentials,
even though often fairly short-range, yield good agreement with experiment.
Moreover, {\it ab initio} TB interactions, which normally include
longer-range interactions, are not affected by this situation.

The first amorphous materials to be studied with TBMD were the elemental
semiconductors silicon and carbon. Silicon, which has a well-defined $sp3$
bonding, is a relatively straightforward choice; TBMD a-Si models are
discussed in the next section. Carbon, on the other hand, is much more
difficult because of the added difficulty associated with its many bonding
states; another article in this special issue is dedicated to {\em a}-C and
we therefore do not discuss it in any detail here but just give, for
completeness, a brief overview of work on this material.

Hydrogenation of amorphous semiconductors reduces significantly the strain
and greatly improves the electronic properties by removing deep states in the
gap. This is particularly so for {\it a}-Si, which requires considerable
amounts of H to achieve device-quality electronic properties. TBMD
simulations have been run to study its effect on structural and electronic
properties, in spite of problems associated with its small mass and high
diffusivity. Likewise, TBMD has been used to investigate compounds (mostly
binaries) --- GeSe$_2$, GaN, GaAs --- where additional complications arise
from partly ionic interactions.


\subsection{Elemental semiconductors}

\subsubsection{Amorphous silicon}

Silicon is the material of reference in the study of amorphous semiconductors
and has been simulated numerically using a variety of techniques. It is
widely accepted as a realisation of Polk's idealised continuous random
network (CRN) \cite{polk71}, which regards the material as a {\em collage} of
randomly-oriented, corner-sharing tetrahedra, thus possessing perfect
coordination. Algorithms have been devised for constructing Polk-type CRN's
on the computer, and these will serve as a reference for simulations based on
total-energy minimisation such as TBMD.

As noted earlier, amorphous samples can be prepared in the computer in many
different ways. Quenching from the melt and annealing is a popular method
because it is akin to the real fabrication process. Quench rates for computer
models (in particular TBMD) are however many orders of magnitude larger than
real ones and the main difficulty therefore resides in cooling slowly enough
for the system to be able to find a reasonable low-energy state. Monte-Carlo
methods such as the Wooten-Winer-Weaire bond-exchange
algorithm\cite{wooten85,wooten87} and the activation-relaxation technique of
Barkema and Mousseau\cite{barkema96,mousseau98}, can yield lower-energy
configurations, but at the expense of a heavier computational effort unless
they are conducted, in part, using empirical potentials.

Kim and Lee\cite{kim94} (KL) have used TBMD (with the GSP model) and the
melt-and-quench approach to construct a 64-atom model for a-Si. The liquid
was produced by running at high temperature (1750 K) for 8 ps. (The liquid
phase itself is much better described by the GSP TB interaction than by the
SW potential; in particular, the number of nearest neighbours and the angular
distributions are in close agreement with the values obtained from {\em ab
initio} MD\cite{virkkunen91,wang92}). The cell was then cooled at a rate
which amounts to $10^{15}$ K/s. The simulations were done in the
microcanonical ensemble at the density of the liquid, about 10\% larger than
that of the crystal, which is itself 1.6\% denser than the amorphous phase
\cite{roorda91}. Thus, the final amorphous phase is under high compressive
strain. The structural properties of the resulting structure are given in
Table \ref{tab:sicoord}, where they are also compared with the predictions of
other models discussed below. The total coordination number, 4.28, is large,
a consequence of the excessive cooling rate but also of the compressive
strain on the system which severly hampers relaxation in view of the short
period of time covered by the simulation. The model displays a rather clean
separation between first- and second-neighbour shells despite an
unrealistically large number of defect states (floating bonds) in the
electronic bandgap, mostly due to the strain associated with
overcoordination.

Servalli and Colombo \cite{servalli93} have studied the influence of the
cooling rate on the structure the material. They considered a 64-atom system,
first ran for 27 ps in the liquid phase, then cooled at rates in the range
$0.096-4\times10^{14}$ K/s --- that is up to 100 times slower than in KL's
simulations. The volume was varied linearly with temperature between values
appropriate for the liquid and amorphous phases. Overall, the coordination is
in much better agreement with experiment than KL's model. Moreover, there
appears to be a strong correlation between the cooling rate and the
``quality'' of the sample. From visual inspection of the data plotted in the
original article, the bond angle distribution, in particular, decreases by
about 20\% between fastest and slowest cooling rate, coming quite close in
the latter case to experimental values. Even the slowest rate, however, might
not be sufficient for proper relaxation: In one run, for instance, a
``major'' relaxation event was observed after about 57 ps, indicating that
the relaxation time is much longer than a few phononic periods even in such
small systems. Servalli and Colombo also prepared a 216-atom {\it a}-Si model
\cite{servalli93} which, because of computational limitations, was
quenched-in over a fairly short time of 11.4 ps. In spite of this, the
electronic density of states is comparable to that obtained from a smaller
system relaxed for a longer time, suggesting that the size of the unit cell
might be an important factor in the total stress found in computer-generated
samples.

In order to minimise the computational workload imposed by the long quenching
process, mixed approaches, using empirical potentials for the time-consuming
preparation followed by TBMD relaxation, can also be used. Mousseau and
Lewis\cite{mousseau97a,mousseau97b}, for example, have combined TBMD with an
efficient optimisation scheme based on classical potentials. A ``rough''
216-atom model was first prepared from a randomly-packed configuration using
the activation-relaxation technique\cite{barkema96,mousseau98} together with
a Stillinger-Weber-type potential. The resulting structure was then relaxed
using TBMD and the GSP potential. The final TBMD stage ensures that the model
is physically realistic; the initial empirical-potential relaxation is
therefore, to some extent, artificial --- as is also the case of the
Wooten-Weiner-Weaire bond-switching process --- but ensures, when combined
with such a powerful optimisation scheme as the activation-relaxation
technique, that the structural model is fully optimised. This is actually
demonstrated in Table \ref{tab:sicoord}: the average coordination number is
very close to four (but slightly below), as expected, and perhaps more
significant, the width of the bond angle distribution is much smaller than
obtained in any other model.

An alternative approach was proposed by Drabold {\em et al.}\cite{drabold90};
it consists in ``incompletely melting'' the sample before cooling. This is
achieved by introducing a vacancy, which destabilizes the lattice: because
the system is small (about 64 atoms in this case), the melting temperature
decreases markedly and relaxation proceeds more easily. After a short
microcanonical run at high temperature, the system is taken down to zero
temperature and relaxed; the calculations were performed using the
Sankey-Drabold ``{\it ab initio} TB'' scheme \cite{sd}. Although this
technique allows for a rapid production of samples, it does not seem at
present to be able to yield quality amorphous networks; the radial
distribution functions (RDF) of the samples prepared in Ref.\
\cite{drabold90} either contain traces of crystallinity after quench or show
the presence of high levels of strain. A more controlled procedure could, in
principle, lead to relatively good structures without requiring full melting
of the crystal.

The atomic structure of {\it a}-Si is not known in detail from experiment and
it is therefore difficult to assess the various computer models. Computer
model preparation is therefore a challenge by itself as much as a necessary
first step for further studies. For these reasons, relatively little effort
has been spent on the actual properties of the material. One exception is the
study by Drabold, Fedders and co-workers of dynamical fluctuations, both
structural and electronic\cite{drabold91}. Using a 63-atom unit cell obtained
using the method discussed above, they followed the electronic states as a
function of time and showed that these could fluctuate significantly even at
350 K\cite{drabold91}. In particular, they observed the lowest unoccupied
molecular orbital to show larger fluctuations than the states at the top of
the valence band. Moreover, localised states fluctuate more than the extended
states. Although the latter observation can be understood simply in
geometrical terms --- the effective mass of the localised state is smaller
than that of the extended state, the former one is not well explained and
requires more detailed characterisation. Studying the structural consequences
of charged defects in their unit cell, Fedders {\it et al.} found that an
electronic transition can induce structural rearrangements that involve up to
many tens of atoms\cite{fedders92a}. The size of the unit cell prevents
quantitative predictions to be made, but these results raise interesting
questions which need to be addressed on larger networks.

Another application is that of De Sandre {\it et al.}\cite{sandre96} who have
computed the elastic constants as a function of temperature for the 64-atom
TBMD model prepared by Servalli and Colombo \cite{servalli93}. At finite
temperature, the elastic constants are defined as the sum of three
contributions: potential, kinetic and fluctuations\cite{sandre96}. Comparison
with experiments and results from the empirical SW model reveals that the TB
potential does better than SW but some discrepancy with experiment remains.
As a general trend, the elastic constants seem to soften as a function of
temperature. The exact relation, however, is somewhat hindered by significant
fluctuations that could be related to slow relaxation processes taking place
during the simulation.

The physical origin and the density of defect states in the bandgap of {\it
a}-Si --- in particular the band tails --- is of both fundamental and
practical interest. Experimentally, it is known that {\it a}-Si contains up
to 1\% of defects \cite{roorda91}, preventing its use in electronic devices.
Direct comparison with experiment requires large unit cells in order to
provide a proper description of the gap region, and thus TBMD cannot be used
for this purpose. Large empirical models can however be used ``as is'' in
order to compute the TB electronic structure, since this requires a single
matrix diagonalisation. This was done by Mercer and Chou\cite{mercer91} as
well as Holender and Morgan\cite{holender92}, who examined 588- and
13824-atom unit cells, respectively. Figure \ref{fig:holender} shows the
electronic density of state for a series of models relaxed with either the
Keating or the Stillinger-Weber potential\cite{holender92}. The density of
states in the gap clearly correlates with the number of coordination defects.
As shown in Fig.\ \ref{fig:holender} (b) (see also Ref.\ \cite{fedders92a}),
there is, however, no one-to-one correlation between coordination and
electronic defects. Some coordination defects yield states which are deep in
the valence band while four-fold coordinated atoms with highly-distorted
environments can form trap states at midgap. The ``rule'' that arises is
therefore that highly-strained networks, even perfectly coordinated, are more
prone to give rise to defects in the electronic gap than low-energy amorphous
structures with even a few coordination defects.

Recently, Dong and Drabold (DD)\cite{dong98} have reported a detailed study
of the band-tail states in an unrelaxed 4096-atom Wooten-Winer-Weaire model
of {\it a}-Si produced by Djorjevi\'c {\it et al.} \cite{djordjevic95}. DD
find that the valence band tail is well described by an exponential decay
$\rho(E) \equiv \exp(-E/E_0)$ with $E_0=190$ meV. These band-tail and gap
states also show a significant degree of spatial localisation (Fig.
\ref{fig:bandtail}). This implies that these states cannot conduct current
unless their density is such that there is a significant overlap between
them. Showing that many localised states actually do overlap significantly
with other states of similar energy, DD proposed a mechanism --- the
``resonant-cluster proliferation'' --- that could lead to conduction by
percolation of overlapping localised states of similar eigen energy through
the whole system, thus explaining the existence of a ``mobility edge'' in
amorphous materials.

\subsubsection{Amorphous carbon}

TB studies of carbon are the object of another article in this Special Issue;
we therefore do not provide here an exhaustive review. For completeness,
however, we mention some relevant work. The modelling of amorphous carbon is
complicated by the many bonding states that carbon can exist in --- $sp$,
$sp2$ and $sp3$ --- and developing a TB potential that can properly account
for the delicate balance between these three states is difficult.
Nevertheless, because of the fundamental and technological importance of the
material, there has been a lot of activity in this field.

The structure of {\it a}-C depends on the method of preparation: material
prepared by evaporation or sputtering tends to be $sp2$ rich while that
produced by mass-selected ion-beam deposition presents large concentration of
$sp3$ (diamond-like) carbon. This situation is also found numerically:
Comparing results for 216-atom unit cells of {\it a}-C prepared at different
densities, Wang {\it et al.} \cite{wang93a} found that the static structure
factor of the low density cell (2.2 g/cm$^3$) gives best agreement with
experimental data for sputtered {\it a}-C\cite{li90}. In this structure,
prepared from the liquid phase by quenching at a rate of $5\times 10^{14}$
K/s, 80.6\% of the atoms are three-fold coordinated, while 7.4 and 12\% have
four and two neighbours, respectively. Wang {\it et al.} further observed
that the large density of three-fold carbons, which form compact regions
surrounded by two- and four-fold coordinated atoms, leads to fully conducting
--- and electronically uninteresting --- materials. Much effort has therefore
been directed towards generating dense, and insulating, tetrahedral amorphous
carbon using different TB models \cite{wang93b,drabold94,lee94,kohler95}, as
we discuss next.

Lee {\it et al.} \cite{lee94} have developed a semi-empirical TBMD model
which appears to provide a reasonable description of the structural
properties of tetrahedral {\it a}-C. However, the complete absence of a gap
in the electronic density of states reveals some inadequacies in the model.
Also using a semi-empirical TBMD model, Wang {\it et al.}\cite{wang93b} have
studied dense {\it a}-C and found results consistent with the predictions
from the more accurate ``{\it ab initio}-type TB'' models of Drabold {\it et
al.}\cite{drabold94} and K\"ohler {\it et al.}\cite{kohler95}. A most
surprising feature of these models is the presence of a single wide band in
the vibrational density of states in the range 300--1400 cm$^{-1}$ that
contrasts markedly with the sharp structures found in both graphite and
diamond. As the density of the material decreases, the band splits into two
wide peaks. Fig.\ \ref{fig:vdos-ac} shows the vibrational density of state
for 128-atom models of {\it a}-C at different densities\cite{kohler95}. It
has been proposed by K\"ohler {\it et al.}\cite{kohler95} that this peak
structure arises from the large strain variations about the C tetrahedra, but
this remains to be established more precisely.


\subsection{Hydrogenated amorphous silicon}

Because it contains a high concentration of defects, which depends strongly
on the mode of preparation, {\it a}-Si usually cannot be used directly in
electronic devices. As already noted in Section \ref{sec:hydrogen}, hydrogen
is often intentionally included in the material so as to saturate the
dangling bonds, thus reducing the concentration of defects to an acceptable
level. Proper models of {\it a}-Si are therefore prerequisite to a detailed
study of {\it a}-Si:H. Describing the interactions of H with Si within a TB
scheme however is a difficult exercise, not only because of the added
complexity of multi-atom interactions, but also because of the subtle
quantum-mechanical bonding properties of hydrogen. Several models have been
developed over the last few years\cite{pan94,bou94,holender93,li94}, based on
either the semi-empirical GSP scheme or the {\em ab initio} Sankey-Drabold
Harris-functional approach (SD-TB)\cite{sd}; the latter defines in a more
natural way the long-range interactions and the complex nature of hydrogen
bonding.

Models can be prepared either by introducing H in existing {\it a}-Si
samples\cite{holender93,fedders93,park95} or by quenching a Si-H mixture from
the melt\cite{tuttle96,lanzavecchia96}. The latter approach was employed by
Tuttle and Adams (TA) \cite{tuttle96} as well as Lanzavecchia and Colombo
(LC) \cite{lanzavecchia96}. Both models contain 216 Si atoms and either 24 or
26 H (i.e., about 10\%). Using the SD-TB scheme, TA first quenched their cell
from 1800 to 300 K at a rate of 10$^{15}$ K/s, equilibrated it for a short
0.5 ps, annealed it at 1200 K for 1.0 ps, then quenched it again to 300 K at
a rate of 10$^{14}$ K/s. The final configuration (at 300 K) presents a
relatively high concentration of defects: 1.5\% of Si atoms are three-fold
coordinated while 9.0\% are five-fold coordinated, which leads to a gapless
electronic density of states. For its part, hydrogen is fully bonded to Si,
and found exclusively in monohydride configurations. The partial radial
distribution functions, shown in Fig.\ \ref{fig:asihrdf}, indicate that the
silicon backbone is essentially identical to that of a pure {\it a}-Si
network and that there is a complete absence of medium-range order in
hydrogen. Most interestingly, the addition of hydrogen is sufficient to
create local inhomogeneities in the cell. A small cavity, for example, is
formed in the cell (Fig. \ref{fig:cavity}), with H atoms ``decorating'' its
internal surface.

A similar calculation was carried out by LC using the GSP-TB model for Si and
a similar one for Si-H and H-H interactions \cite{pan94}, as discussed in
Section \ref{sec:hydrogen}. The cooling rate in this case is four times
smaller than that used by TA. Here again, monohydride complexes are found to
be the most likely configuration for hydrogen. However, LC also found some
SiH$_2$ complexes as well as small three- and four-atom H clusters. In spite
of their large density of structural and electronic defects, these two models
provide important hints on the structural modifications of the amorphous
phase which might be induced by H cavities, hydrogen chains and clusters. The
difference between TA and LC structures, in terms of the presence or absence
of cavities and polyhydrides complexes, is not understood at the moment. It
might be due to different modes of preparation or potentials, or to the fact
that TA used a rescaled mass for H while LC employed the actual value.

Starting from a ``preformed'' {\it a}-Si sample offers the advantage that H
can be introduced in a controlled manner and the release of electronic and
elastic strain monitored, in particular through the removal of electronic
traps. Gap states do not arise only from the presence of dangling bonds but
also in strained environments, and hydrogen must therefore sometimes be
forced in by breaking the network. For example, Fedders and
Drabold~\cite{fedders93} and Holender {\it et al.}\cite{holender93} simply
removed overconstrained silicon atoms and saturated the newly-formed dangly
bonds with hydrogen. Evidently, such a procedure has little in common with
the actual physical process, but nevertheless proves useful insights on how
hydrogen releases the strain and causes states to move deeper in the valence
band.

The Staebler-Wronski effect\cite{staebler77}, i.e., the degradation of the
photoelectric conversion properties of the material under exposure to light,
is the motivation behind much of the work on {\it a}-Si:H. This phenomenon is
known to involve structural rearrangements following the absorption of
photons, and can in fact be reversed by annealing at sufficiently high
temperature. It is however not clear if the electron emitted following the
absorption of the photon plays an active role or if it merely transfers
kinetic energy to the network. A variety of approaches have been used, within
the TBMD framework, to investigate this problem. Using a 62-Si plus 5 or 7-H
atom cell, Fedders\cite{fedders95} has examined the effect, on energy and
relaxation, of the state of charge of dangling bonds and finds the
passivation energy to strongly depend on the local environment, fluctuating
by as much as 1 eV in their limited sample. It also appears that charged
dangling bonds would be more energetically favourable than neutral ones. It
is difficult experimentally to establish the density of specific charge
defects so that more numerical work is needed to confirm and refine the
results of these calculations.

A similar study was carried out by Biswas {\it et al.} (BLYB) \cite{biswas97}
using the 60-atom (54 Si + 6 H) model of Guttman and Fong\cite{guttman82}.
Here the cell was more completely relaxed by equilibrating over several tens
of ps at 300 K while, in contrast, Fedders\cite{fedders95} worked with
configurations optimised locally at 0 K. Adding or removing a H atom, BLYB
followed the relaxation of the lattice. This relaxation takes place very
rapidly, over a period of about 10 ps, and involves almost exclusively the Si
atom bonded to the defect, with very similar behaviour for positively and
negatively charged defects. Park and Myles (PM) have proposed a ``hot-spot''
method for stimulating relaxation \cite{park95}. Starting with a 0 K
configuration, two atoms on a bond are given a burst of energy (2 eV in this
case), and the dynamics is followed until the excitation reaches the boundary
of the periodic box (300 fs). Using the 60-atom Guttman-Fong model, Park and
Myles have found, of all the bonds they tried, only one for which the
hot-spot excitation leads to a new configuration, producing a dangling and a
floating bonds, after jumping an activation barrier of about 2 eV. This is in
agreement with BLYB's results where the localised and rapid relaxation is
also an indication of a very stable configuration. In the context of the
Staebler-Wronski effect, these results would support a very localised
mechanism controlled by isolated bonds being broken and reforming. However,
the stability of the lattice could also be due the its very small size; more
simulations are needed in order to establish the microscopic origin of this
effect.

Other dynamical properties of {\it a}-Si:H have also been studied. Fedders
and Drabold (FD), for instance, have searched their model for conformational
fluctuations by quenching snapshot configurations at regular intervals in
time during a 600 K run\cite{fedders96}. They found that although the
connectivity of the lattice is preserved, the quenched configurations are all
slightly different, revealing the presence of a large number of very small
barriers between slightly different states. The most relevant barrier is that
for the diffusion of H. This light atom is likely to move significantly in
the mostly-empty structure. However, as was the case for {\em c}-Si, the
nature of the diffusion mechanism remains unclear. From simulations on a
system containing 216 Si and 24 H atoms, Lanzavecchia and Colombo found that
H migrates through jumps between neighbouring dangling bonds {\em via} a
metastable Si-H-Si state\cite{lanzavecchia96}. Similar observations were
reported by Tuttle and Adams \cite{tuttle97} and Biswas {\it et
al.}\cite{biswas98}: during diffusion, a hydrogen binds to an already
four-fold-coordinated atom, giving rise, for a short period of time, to a
floating bond; Biswas {\it et al.} estimate the formation energy of this
floating bond to be in the range 1.3--2.3 eV.


\subsection{Compounds}

Amorphous compounds, such as the III-V semiconductors and the chalcogenides,
are particularly challenging for simulations. First is the problem of
constructing an appropriate set of interactions: not only are there more
interaction types than in mono-atomic compounds, but also ionisation and
charge transfer effects can become important, requiring special attention.
Second, the potential energy surface is complicated by the introduction of a
new dimension, namely the chemical identity of the constituent species. Thus,
in a study of the liquid-to-amorphous transformation, for instance, the
search for the ground state must seek to minimise both the topological and
the chemical disorder. As a consequence, the relaxation timescales available
through TBMD simulations, which are already out of measure with experimental
timescales, become prohibitively long, except perhaps for some very ionic
compounds, such as SiO$_2$, where even the liquid phase already exhibits
almost perfect chemical order.

In spite of this difficulty, most TBMD studies of amorphous III-V materials
proceeds via the usual melt-and-quench cycle. Indeed, there exists no
``recipe'' {\em \`a la } Wooten-Winer-Weaire to prepare amorphous compounds
that are chemically ordered. Further, there exists no satisfactory classical
potentials for these materials that can be used to bypass the
computer-intensive TBMD melt-and-quench cycle. Nevertheless it is possible,
through a combination of approaches, to generate very high-quality models, as
will see below.

TBMD can provide important insights on the physics of these materials in
spite of the above limitations. This is demonstrated, for instance, in the
recent simulation by Stumm and Drabold of GaN \cite{stumm97}. Using the
Sankey-Drabold TBMD model\cite{sd}, Stumm and Drabold quenched a 64-atom cell
of liquid GaN into the amorphous phase at a rate of $1.3\times10^{15}$ K/s.
Two densities were studied: one equal to that of wurtzite GaN and the other
at 82\% of this density. For both structures, the total radial distribution
function exhibits a deep minimum between the first and second-neighbour peaks
despite a large concentration of three-fold atoms (37 and 66\%,
respectively). Contrary to what is seen in {\it a}-C, where a large density
of three-fold atoms fills the electronic gap, there are no states which
appear in the gap of either GaN unit cells. Amorphous GaN is thus predicted
to be, in its own right, an interesting material for device applications.

Studies of {\it a}-GaAs obtained by quenching from the liquid phase have been
carried out by Molteni, Colombo and Miglio (MCM) \cite{mol93}, and Seong
and Lewis (SL) \cite{seong96}. This material has a relatively small ionicity;
the TB-potential proposed by Molteni {\it et al.} \cite{molteni94a}, already
discussed in Section \ref{sec:gaas}, deals with charge-transfer effects by
introducing a screened electrostatic term, thus keeping the potential short
range. In both cases, 64-atom cells were used and the cooling rates were
similar ($1.5\times10^{12}$ K/s for MCM, 50\% slower for SL). One important
difference however is in the choice of density: while MCM fixed it at the
crystalline value, SL chose to optimise it; a density 3.2\% smaller than the
crystalline one was thus obtained (Fig. \ref{fig:density}), in agreement with
experiment --- values in the range 4.98--5.11 g/cm$^3$ have been reported;
the crystalline value is 5.32 g/cm$^3$. Optimisation of the density also
leads to significant, and perhaps unexpected, differences between the two
structures: MCM find the bonding environments of As and Ga to be symmetric,
that is the two species exhibit similar distributions of coordination
defects; in contrast, SL observe sizable differences between the two species,
with Ga much more likely to be in a five-fold coordination state than As,
which prefers to be in a three-fold state. Overall, however, the average
coordination is almost precisely four --- in fact slightly less, 3.94,
certainly related to the fact that {\it a}-GaAs is less dense than {\it
c}-GaAs.

TBMD has also been used to generate models for the chalcogenide glasses. Cobb
and co-workers \cite{cappelletti95,cobb96}, for instance, have quenched 62-
and 216-atom models of GeSe$_2$ from the liquid phase using the
Sankey-Drabold-TB scheme \cite{sd}. The total simulation time for the
216-atom cell is 5 ps. In spite of this very short time, which leaves a large
number of defects in the network, the system exhibits a well-defined optical
gap of 1.72 eV, free of defect states; the static structure factor and
vibrational density of states is also in reasonable agreement with experiment
(Fig. \ref{fig:vdos}). Most ($\sim$85\%) Ge atoms are fourfold coordinated
with about 26\% Ge having another Ge atom as one of its neighbours; a similar
fraction of Se form homopolar bonds. Because of different coordination, this
leads to a total density of ``wrong'' bonds of 10.8\%. Cobb {\it et al.},
furthermore, find that the localised electronic states are far from the band
edges, thus leaving a wide, state-free gap.

Because GeSe$_2$ is significantly less ionic than, e.g., SiO$_2$, the
timescale needed to obtain a chemically-ordered GeSe$_2$ glass is well beyond
the reach of TBMD. This is true also of the III-V compound {\em a}-GaAs. As a
result, the melt-and-quench simulations of {\em a}-GaAs mentioned above lead
to a density of defects which is large. Wrong bonds, for instance, are found
to be in concentrations of 12.2 and 12.9\% in the models of SL and MCM,
respectively. While the exact number is not known from experiment, these
values are most certainly on the high side. The full TBMD melt-and-quench
cycle can be partly bypassed through a combination of approaches. As noted
already, there exists not equivalent, in the case of III-V compounds, to the
Wooten-Winer-Weaire model for {\em a}-Si; however, even though this model is
``non-physical'' (the topology is distorted in an {\em ad hoc} manner), it
nevertheless yields structures in excellent agreement with experiment. In the
same spirit --- i.e., the end justifies the means --- Mousseau and Lewis
\cite{mousseau97a,mousseau97b} have proposed a scheme for {\em a}-GaAs based
on empirical potentials that compares favourably with melt-quenched models,
as we discuss now.

Amorphous covalently-bonded semiconductors have traditionally been described
in terms of CRN's. Polk's CRN \cite{polk71}, presumably appropriate to {\em
a}-Si, consists of a {\em collage} of corner-sharing tetrahedra which
preserves the ideal coordination of four. The Connell-Temkin CRN
\cite{connell74} is similar, except for the additional constraint that
odd-membered rings are not permitted. This model is likely relevant to
chemically-ordered binary compounds since the presence of odd-membered rings
necessarily imply the presence of wrong-bond defects. That {\em a}-GaAs is
indeed akin to the Connell-Temkin model has been recently established by
Mousseau and Lewis \cite{mousseau97a,mousseau97b}. This was done using a
combination of approaches as follows: Starting with a 216-atom supercell in a
random state, ``zeroth-order'' models were generated using Barkema and
Mousseau's activation-relaxation technique (ART), the relaxation being
carried out using a simple Stillinger-Weber-like potential with and without
an additional term between like atoms; these structures were then relaxed
using the MCM-TBMD model. Mousseau and Lewis have found the
Connell-Temkin-type network to have significantly lower energy than the
Polk-type network and therefore to provide a better representation of {\em
a}-GaAs as can be seen in Tables \ref{tab:coord} and \ref{tab:energies}.
However, the same structures with Ga and As replaced by Si and relaxed using
the GSP-TBMD showed no difference in energy, indicating that both models have
the same inherent strain level. This last result is rather surprising: the
density of odd-membered rings is apparently not determined by elastic
constraints but, simply, by entropic considerations.

Following the static minimisation stage, both GaAs models were relaxed at
finite temperature (300 K for 7 ps and 700 K for 8.8 ps) to assess their
stability and compute their dynamical properties. It was found that the Polk
CRN for GaAs is unstable and in fact distorts significantly at temperatures
as low as 700 K while the Connell-Temkin CRN remains relatively unaffected by
the annealing process. However, a detailed comparison of the structural,
vibrational and electronic properties of the two networks reveals little
qualitative differences between them, indicating that intermediate-range
order plays a relatively minor role in determining the properties of GaAs. It
appears that only direct measurements of the density of wrong bonds can shed
light on the experimental nature of the structure of {\it a}-GaAs.


\subsection{Surfaces of amorphous semiconductors}

Surfaces clearly play an important role in the microelectronic properties of
devices and in particular influence the growth process. However, there has
been relatively few simulations of disordered-material surfaces (including
growth) in the TB framework, possibly because these require, to start with,
realistic (and large enough) sample of the bulk amorphous phase. It is
important to note that the timescale for growth by atom deposition is much
shorter in computer models than in real experiments, so that it is difficult
to extract quantitative information from such simulations.

Kilian {\it et al.}\cite{kilian93} have used the {\it ab initio} TB scheme of
Sankey and Drabold\cite{sd} to study the surface states of {\it a}-Si and
{\it a}-Si:H. A 216-atom Wooten-Weiner-Weaire model of {\em a}-Si
\cite{wooten85,wooten87} was first relaxed using TBMD. The periodic boundary
condition was then removed along the $z$ direction and the bottom layer
passivated with hydrogen; the top layer, finally, was relaxed with and
without hydrogen in order to study localised electronic surface states and
the effect of hydrogenation on them. In order to do this, a local charge
$q(n,E)$ for atom $n$ is defined, where E is the energy eigenstate. One may
then calculate the mean square charge of a ``layer'' located at position $z$
and containing $N_z$ atoms:
\[
   q_2(E,z) = \sum_{n_z}^{N_z} q(n_z,E)^2.
\]
Multiplying this quantity by $N_z$ gives a measure of the localisation of
charge in the layer, $Q_2(E,z)$; in this way, the decay of surface states
with depth can be monitored. Figs.\ \ref{fig:sketches} and \ref{fig:Q2H} show
some surface defects --- not necessarily topological -- and the corresponding
spatial localisation of charge. Hydrogenating the surface does not eliminate
these states, however, contrary to what often happens in the bulk: Fig.\
\ref{fig:Q2H} shows that surface states are more localised by the addition of
hydrogen but not otherwise affected. The study of other {\it a}-Si:H models
\cite{kilian97} lead to similar conclusions: surface states seem to be
difficult to eliminate. It should be noted however that, again because of
computational limitations, the surfaces may not be fully relaxed. It remains
to be seen if detailed thermal annealing would decrease the influence of
localised surface states.


\section{Concluding remarks}

Tight-binding molecular dynamics has contributed significantly to our
understanding of defects in semiconductors. Because it is generally reliable,
and yet economical from the computational viewpoint, the method constitutes
an excellent complement to {\it ab initio} calculations. While the latter is
limited to small unit cells, for which self-interaction of defects remains
important, it is possible with TBMD to deal with systems containing many
hundreds of atoms in a fully-relaxed configuration. With the advent of ${\cal
O}(N)$ methods, it becomes feasible to carry out TBMD studies of more complex
defects involving many tens of atoms, a task which is at present beyond the
reach of {\it ab initio} approaches.

The study of amorphous semiconductors has also greatly benefited from TBMD
calculations. With the wide range of local environments found in these
materials, it is difficult to construct a fully satisfactory empirical
potential; TB models have proved much more transferable. However, because of
the computational cost of TB calculations, still much larger than classical
potentials, and the difficulty of implementing ${\cal O}(N)$ methods for
these materials, simulations have been limited to, principally, 64-atom cells
with some noteworthy exceptions up to 512-atom cells. Using a mixture of
hands-on or empirical approaches and TBMD for creating the initial structures
can compensate for the increased computational cost. In spite of the biases
that they possibly introduce, such mixed approaches appear to be the best, at
present, for studying large, low-strain disordered structures.

The examples presented in this article underline the important contributions
of TBMD calculations to the study of semiconductors. As the method becomes
more widespread and fundamentals are established, it can be expected that
several of the questions left unanswered in this review will be addressed in
the coming years.


\section{Acknowledgements}

We are grateful the many people who have aroused our interest in this field
and who have helped us in one way or another, in particular L. Colombo, E.
Hernandez, I. Kwon, and H. Seong. This work was supported by grants from the
Natural Sciences and Engineering Research Council (NSERC) of Canada and the
``Fonds pour la formation de chercheurs et l'aide {\`a} la recherche'' of the
Province of Qu{\'e}bec.


\bibliographystyle{prsty}

\newpage
\widetext
\onecolumn

\begin{table}[t]
\caption{
Point defect formation energies (in eV) obtained using various TB models and
comparison with the LDA results. The relaxation energies are given in
parenthesis (when available). The LDA results for the two interstitials are
not consistently relaxed because of the use of different computational
schemes (e.g., supercell versus Green's functions). For the monovacancy and
the divacancies, the LDA calculations allowed full relaxation. MC: Mercer and
Chou \protect\cite{mercer91}; KBWHS: Kwon {\em et al.}\protect\cite{kwo94};
LKKVERYA: Lenosky {\em et al.}\protect\cite{len97}.
}
\begin{tabular}{l|ccccccccc}
                  & GSP         & GSP         & GSP         & MC     & KBHS        & KBHS        & LKKVERYA    & LKKVERYA    & LDA                   \\ \hline
                  & $N=64$      & $N=216$     & $N=512$     & $N=64$ & $N=64$      & $N=216$     & $N=64$      & $N=216$     & $N=64$                \\ \hline
References        & \protect\cite{wan91,son93} & \protect\cite{wan91} & \protect\cite{wan91} & \protect\cite{mercer91} & \protect\cite{kwo94} & \protect\cite{kwo94} &
    \protect\cite{len97} & \protect\cite{len97} & \protect\cite{bar84,kel92,seo96} \\ \hline
Defect            &             &             &             &        &             &             &             &             &                       \\ \hline
Monoacancy        & 3.67 (1.35) & 3.96 (1.86) & 4.12 (1.91) & 3.76   & 3.46 (1.26) & 3.93 (1.64) & 3.40 (0.24) & 3.78 (0.20) & 3.3--4.3 (0.4--0.6)   \\
T interstitial    & 4.39 (2.13) & 4.40 (2.14) & 4.41 (2.14) & 4.95   & 3.61 (0.51) & 4.42 (0.49) & 3.55 (0.59) &             & 3.7--4.8 (0.1--0.2)   \\
H interstitial    & 5.78 (4.30) & 5.90 (4.42) & 5.93 (4.42) &        & 4.75 (1.17) & 5.13 (1.23) & 3.56 (0.89) &             & 4.3--5.0 (0.6--1.1)   \\
Divacancy         & 5.68 (1.73) &             &             &        &             &             &             &             & 4.32 (0.27)           \\
Split divacancy   & 6.54 (2.43) &             &             &        &             &             &             &             & 5.90 (0.57)           \\
Frenkel pair      & 6.55 (3.89) &             &             &        &             &             &             &             &                       \\
\end{tabular}
\label{sidfe}
\end{table}

\begin{table}[t]
\caption{
Energy levels of various neutral defects in their relaxed configurations, in
eV, measured with respect to the valence-band maximum. GSP-TBMD: Results from
Ref.\ \protect\cite{son93}; LDA: Results from Ref.\ \protect\cite{seo96}.
}
\begin{tabular}{l|ccc}
Defect          & GSP-TBMD         & LDA  \\ \hline
Monoacancy      & 0.76             & 0.23 \\
T interstitial  & 0.52, 0.12       &      \\
Divacancy       & 1.00, 0.46       & 0.04 \\
Split divacancy & 0.76, 0.46, 0.34 & 0.08 \\
Frenkel pair    & 0.76, 0.52       &      \\
\end{tabular}
\label{sideflev}
\end{table}

\begin{table}[t]
\caption{
$s$-like ($A_1$ symmetry) deep energy levels in GaP and Si, measured with
respect to the top of the valence band; C.B.\ indicates levels lying in the
conduction band. TB: TB model of Hjalmarson {\em et al.}\protect\cite{hja80},
which does not include relaxation. TBMD: TB model of Hjalmarson {\em et
al.}\protect\cite{hja80} including relaxation, after Li and Myles, Ref.\
\protect\cite{lim91}, where experimental references can also be found. The
equilibrium bond lengths are 2.36 and 2.35 \AA\ for GaP and Si, respectively.
}
\begin{tabular}{l|cccc}
System           &  Bond length & TB   & TBMD  & Experiment  \\ \hline
                 & \AA\         & eV   & eV    & eV          \\
GaP:N            & 2.15         & 2.10 & 2.25  & 2.34        \\
GaP:O            & 2.78         & 1.85 & 1.70  & 1.46        \\
GaP:P$_{\rm Ga}$ & 2.28         & 1.03 & 1.09  & 1.10        \\
GaP:Ge           & 2.24         & 1.85 & 1.95  & 2.16        \\
GaP:Se           & 2.22         & 2.32 & C.B.  &             \\
Si:S             & 2.23         & 0.58 & 0.63  & 0.85        \\
Si:Se            & 2.03         & 0.65 & 0.83  & 0.86        \\
Si:Te            & 2.60         & 1.12 & 1.05  & 1.01        \\
Si:C             & 1.98         & 1.09 & C.B.  &             \\
\end{tabular}
\label{sigap}
\end{table}

\begin{table}
\caption{
Nearest-neighbour breathing-mode displacements and local volume changes for
vacancies and antisites in GaAs, after Seong and Lewis\protect\cite{seo95}.
The average relaxations are given both in \AA\ and relative to the bond
distance of bulk GaAs (2.45 \AA). $+$ and $-$ refer to outward and inward
relaxation, respectively. $\Delta V = V-V_0$ is the change in volume of the
defect (i.e., the volume of the tetrahedron formed by the four nearest
neighbours) resulting from relaxation.
}
\begin{tabular}{l|cccc|c|c|c}
Defect & \multicolumn{4}{c} {Breathing (\AA)} \vline & Average (\AA)
       & Average (\%) & $\Delta V/V_0 (\%)$ \\
       & 1 & 2 & 3 & 4 &     &    &    \\ \tableline
$V_{\rm Ga}^+$      & -0.31 & -0.35 & -0.31 & -0.31 & -0.32 & -13.1 & -34.3 \\
$V_{\rm Ga}^0$      & -0.33 & -0.33 & -0.31 & -0.30 & -0.32 & -13.0 & -34.1 \\
$V_{\rm Ga}^-$      & -0.32 & -0.32 & -0.32 & -0.29 & -0.32 & -12.9 & -33.9 \\
$V_{\rm Ga}^{2-}$   & -0.32 & -0.32 & -0.32 & -0.30 & -0.31 & -12.8 & -33.7 \\
$V_{\rm Ga}^{3-}$   & -0.31 & -0.31 & -0.31 & -0.31 & -0.31 & -12.8 & -33.7 \\
\tableline
$V_{\rm As}^{+}$    &  0.38 &  0.38 &  0.38 &  0.38 &  0.38 &  15.5 &  54.0 \\
$V_{\rm As}^0$      &  0.44 &  0.44 &  0.44 & -0.49 &  0.21 &   8.5 &  24.2 \\
$V_{\rm As}^-$      &  0.39 &  0.39 &  0.39 & -0.50 &  0.17 &   6.8 &  18.9 \\
$V_{\rm As}^{2-}$   &  0.45 & -0.42 &  0.45 & -0.42 &  0.02 &   0.8 &  -3.3 \\
\tableline
Ga$_{\rm As}^{2+}$  &  0.49 &  0.49 &  0.21 &  0.21 &  0.35 &  14.4 &  50.0 \\
Ga$_{\rm As}^{+}$   &  0.48 &  0.17 &  0.13 &  0.13 &  0.23 &   9.2 &  30.0 \\
Ga$_{\rm As}^0$     &  0.48 &  0.07 &  0.07 &  0.07 &  0.17 &   7.0 &  22.2 \\
Ga$_{\rm As}^-$     &  0.11 &  0.02 &  0.02 &  0.02 &  0.04 &   1.8 &   5.5 \\
Ga$_{\rm As}^{2-}$  & -0.01 & -0.01 & -0.01 & -0.01 & -0.01 &  -0.3 &  -1.0 \\
\tableline
As$_{\rm Ga}^{2+}$  &  0.06 &  0.06 &  0.06 &  0.06 &  0.06 &   2.5 &   7.6 \\
As$_{\rm Ga}^{+}$   &  0.06 &  0.06 &  0.06 &  0.06 &  0.06 &   2.4 &   7.5 \\
As$_{\rm Ga}^{0}$   &  0.06 &  0.06 &  0.06 &  0.06 &  0.06 &   2.4 &   7.5 \\
As$_{\rm Ga}^{-}$   &  0.06 &  0.06 &  0.06 &  0.06 &  0.06 &   2.5 &   7.8 \\
\end{tabular}
\label{rlxb}
\end{table}

\begin{table}
\caption{
Nearest-neighbour pairing-mode displacements for As vacancies and Ga
antisites, after Seong and Lewis\protect\cite{seo95}.
}
\begin{tabular}{l|cccc|cccc}
Defect & \multicolumn{4}{c} {Pairing 1 (\AA)} \vline &
\multicolumn{4}{c} {Pairing 2 (\AA)} \\
                      &     1 &     2 &     3 &     4 &     1 &     2 &     3 &     4 \\
\tableline
$V_{\rm As}^{+}$   &     0 &     0 &     0 &     0 &     0 &     0 &     0 &     0 \\
$V_{\rm As}^0$     & -0.01 & -0.01 &  0.03 &     0 &  0.02 &  0.02 &     0 &     0 \\
$V_{\rm As}^-$     & -0.01 & -0.01 &  0.02 &     0 &     0 &     0 &     0 &     0 \\
$V_{\rm As}^{2-}$  &  0.01 & -0.01 &  0.01 & -0.01 & -0.01 &  0.01 & -0.01 &  0.01 \\
\tableline
Ga$_{\rm As}^{2+}$ & -0.02 & -0.02 & -0.21 & -0.21 &     0 &     0 &     0 &     0 \\
Ga$_{\rm As}^{+}$  &  0.03 & -0.01 & -0.12 & -0.12 &     0 &     0 & -0.03 &  0.03 \\
Ga$_{\rm As}^0$    &     0 &  0.08 & -0.04 & -0.04 &     0 &     0 & -0.06 &  0.06 \\
Ga$_{\rm As}^-$    &     0 &  0.05 & -0.03 & -0.03 &     0 &     0 & -0.05 &  0.05 \\
Ga$_{\rm As}^{2-}$ &     0 &     0 &     0 &     0 &     0 &     0 &     0 &     0 \\
\end{tabular}
\label{rlx}
\end{table}

\begin{table}
\caption{
Coordination and bond angle distribution for a variety of {\it a}-Si models.
CP: Stich {\it et al.} ({\em ab initio} quench) \protect\cite{stich91}; ML:
Mousseau and Lewis (empirical construction, TB
relaxation)\protect\cite{mousseau97a}; KL: Kim and Lee (TB quench)
\protect\cite{kim94}; DFSD: Drabold {\it et al.} (TB quench)
\protect\cite{drabold90}; SC: Servalli and Colombo (TB quench)
\protect\cite{servalli93}.
}
\begin{tabular}{cllllll}
        &CP     &ML     &KL     &DFSD   & SCA   &SCB    \\ \hline
$Z=3$   & 0.002 & 0.032 & 0.032 & 0.063 & 0.049 & 0.009 \\
$Z=4$   & 0.966 & 0.954 & 0.828 & 0.877 & 0.904 & 0.958 \\
$Z=5$   & 0.032 & 0.014 & 0.125 & 0.060 & 0.047 & 0.033 \\
$Z=6$   & 0     & 0     & 0.016 & 0     &       &       \\
$<$Z$>$ & 4.03  & 3.98  & 4.28  & 4.00  & 3.99  & 4.02  \\
$r_{C}$& 3.0   & 3.0   & 2.75  & 2.70  & 2.82  & 2.82  \\
$\theta$& 106.7 & 109.4 & 108.3 &       &       &       \\
$\Delta
\theta$ &  16.3 &  9.4  &  15.5 &       &       &       \\
\# atoms&  64   & 216   &  64   & 63    &  64   &  64   \\
\end{tabular}
\label{tab:sicoord}
\end{table}
 

\begin{table}
\caption{
Structural characteristics of various models for {\it a}-GaAs. Distribution
of coordination numbers, $Z$ (and nearest-neighbour cutoff distance,
$r_{NN}$), first nearest-neighbour distance, $r_1$ (and width, $\Delta r_1$),
density of wrong bonds, and width of the bond-angle distribution,
$\Delta\theta$. CRN-P: Polk-type continuous random network of Mousseau and
Lewis \protect\cite{mousseau97a,mousseau97b}; CRN-CT: Connell-Temkin-type
continuous random network of Mousseau and Lewis
\protect\cite{mousseau97a,mousseau97b}; SL --- TB simulations of Ref.\
\protect\cite{seong96}; MCM --- TB simulations of Ref.\
\protect\cite{mol93}; CP --- Car-Parrinello simulations of Ref.\
\protect\cite{fois92}.
}
\begin{tabular}{l|ccccccc}
                      & \multicolumn{2}{c}{CRN-P} & \multicolumn{2}{c}{CRN-CT} & SL & MCM & CP \\
                      & 0 K   & 300 K & 0 K   & 300 K & 0 K   & 0 K  & 0 K   \\ \hline
$Z=$ 3                & 0.046 & 0.128 & 0.051 & 0.118 & 0.242 & 0.14 & 0.219 \\
$Z=$ 4                & 0.954 & 0.845 & 0.944 & 0.830 & 0.598 & 0.66 & 0.781 \\
$Z=$ 5                & 0     & 0.026 & 0.005 & 0.045 & 0.129 & 0.18 & 0     \\
$Z=$ 6                & 0     & 0.001 & 0     & 0.004 & 0.024 &      & 0     \\
$Z=$ 7                & 0     & 0.000 & 0     & 0.002 & 0.007 &      & 0     \\
$<Z>$                 & 3.95  & 3.90  & 3.95  & 3.95  & 3.94  & 4.09 & 3.83  \\
$r_{NN}$ (\AA)        & 3.0   & 3.1   & 3.0   & 3.1   & 3.0   & 3.0  & 2.8   \\
$r_1$ (\AA)           & 2.508 & 2.505 & 2.517 & 2.507 &       &      &       \\
$\Delta r_1$ (\AA)    & 0.075 & 0.117 & 0.073 & 0.103 &       &      &       \\
Wrong bonds (\%)      & 14.1  & 14.2  & 3.9   & 5.2   & 12.2  & 12.9 & 10.0  \\
$\Delta\theta$ (deg.) & 11.0  & 14.1  & 10.8  & 15.0  & 17.0  & 17.0 &       \\
\end{tabular}
\label{tab:coord}
\end{table}

\begin{table}[tb]
\caption{
Energy (eV/atom) of the Polk-type (CRN-P) and Connell-Temkin-type (CRN-CT)
models for {\em a-}Si and {\em a-}GaAs continuous random networks of Mousseau
and Lewis, relaxed with TB potentials at 0 K
\protect\cite{mousseau97a,mousseau97b}. For GaAs, we also give the results
from the TB-MD simulations of Seong and Lewis (SL), Ref.\
\protect\cite{seong96}.
}
\begin{tabular}{lccc}
Network &   Si    &  GaAs   &   GaAs (SL) \\ \hline
 CRN-P  & -13.172 & -13.450 &             \\
 CRN-CT & -13.163 & -13.561 &  -13.450    \\
Crystal & -13.389 & -13.802 &  -13.802    \\
\end{tabular}
\label{tab:energies}
\end{table}

\narrowtext
\onecolumn

\newpage
\epsfxsize=10cm
\epsfbox{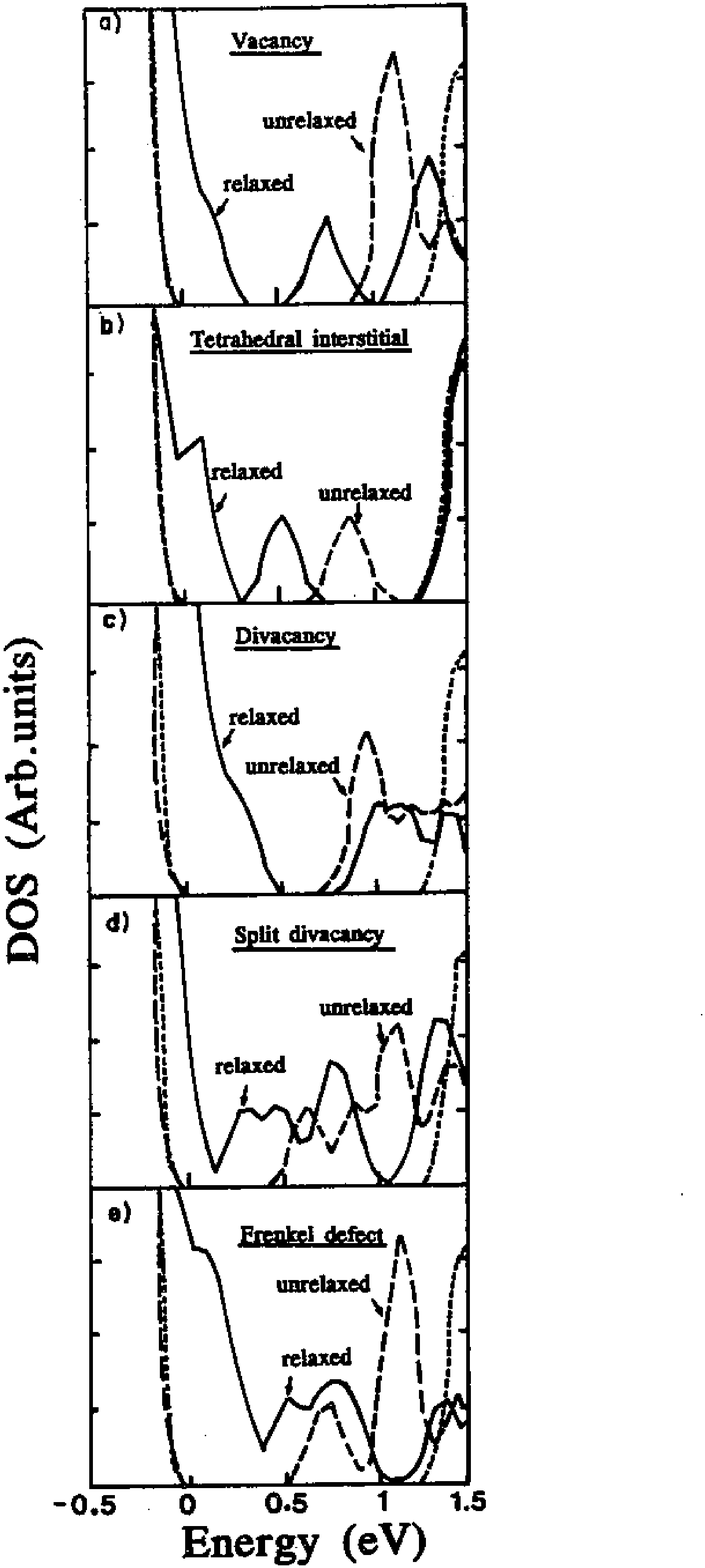}
\vspace{1cm}
\begin{figure}
\caption{
Density of states near the band gap for various defects in crystalline Si.
(From Ref.\ \protect\cite{son93}; reproduced by kind permission).}
\label{sidefdos}
\end{figure}

\newpage
\epsfxsize=10cm
\epsfbox{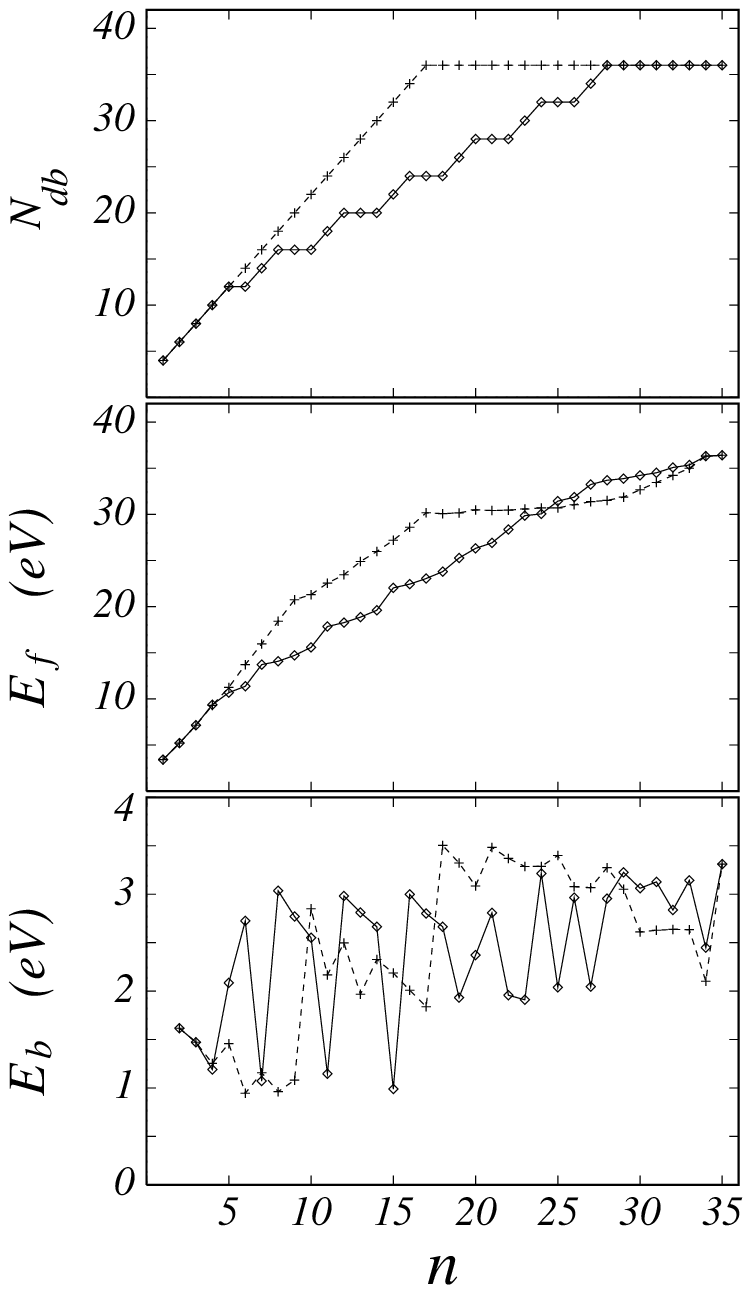}
\vspace{1cm}
\begin{figure}
\caption{
Number of dangling bonds $N_{\rm db}$ (top panel), formation energy $E_f$
(middle panel) and binding energy $E_b$ (bottom panel) for clusters of
vacancies in Si as a function of size $n$. The full lines and diamond symbols
are for HRC clusters, while dashed lines and plus symbols are for SPC
clusters. (From Ref.\ \protect\cite{bon98}; reproduced by kind permission).}
\label{sivaclus}
\end{figure}

\newpage
\epsfxsize=15cm
\epsfbox{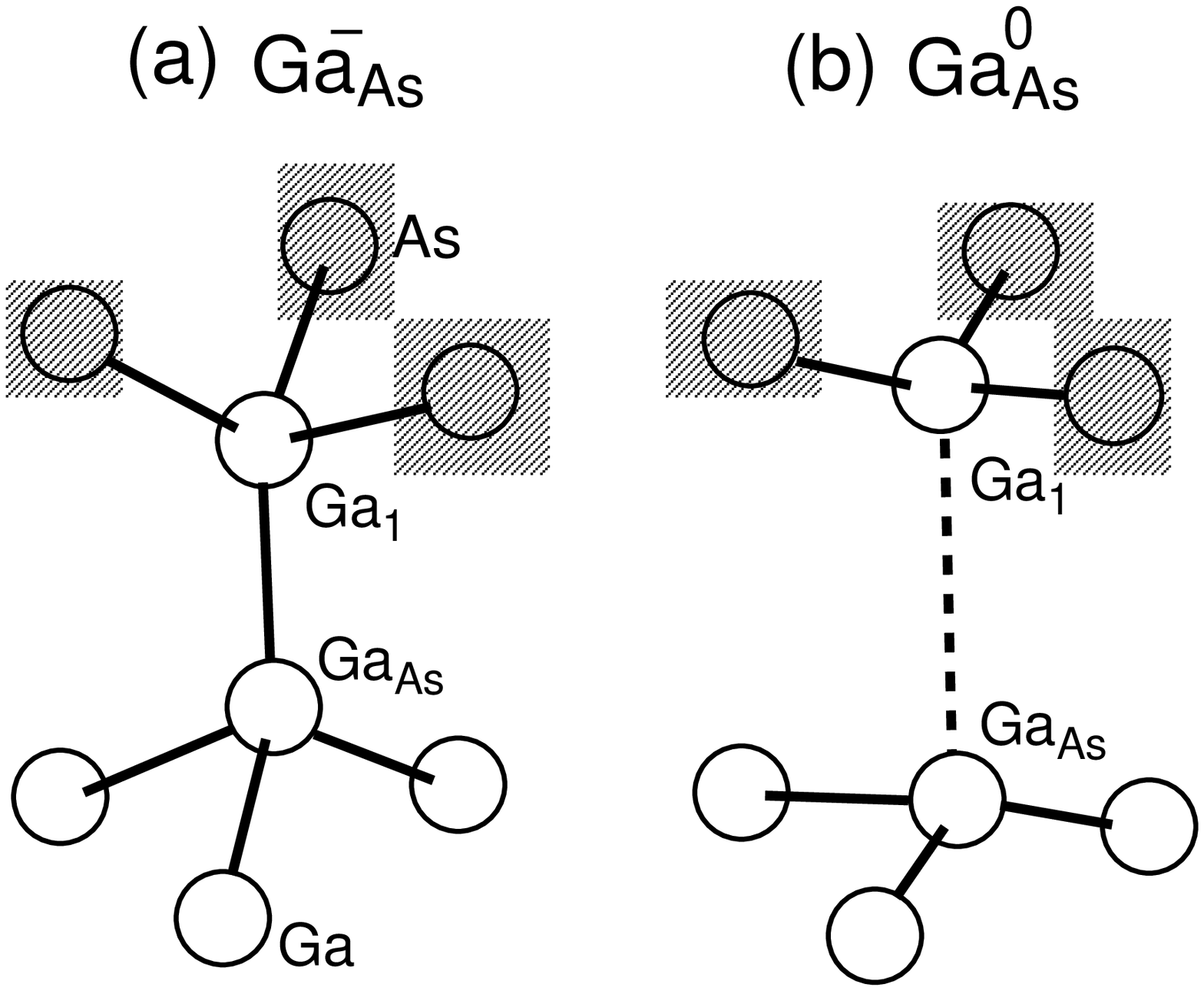}
\vspace{-5cm}
\begin{figure}
\caption{
Local relaxed atomic configuration for (a) the negative and (b) the neutral
Ga antisite in GaAs. (From Ref.\ \protect\cite{seo95}; reproduced by kind
permission).}
\label{ga_antisite}
\end{figure}

\newpage
\epsfxsize=15cm
\epsfbox{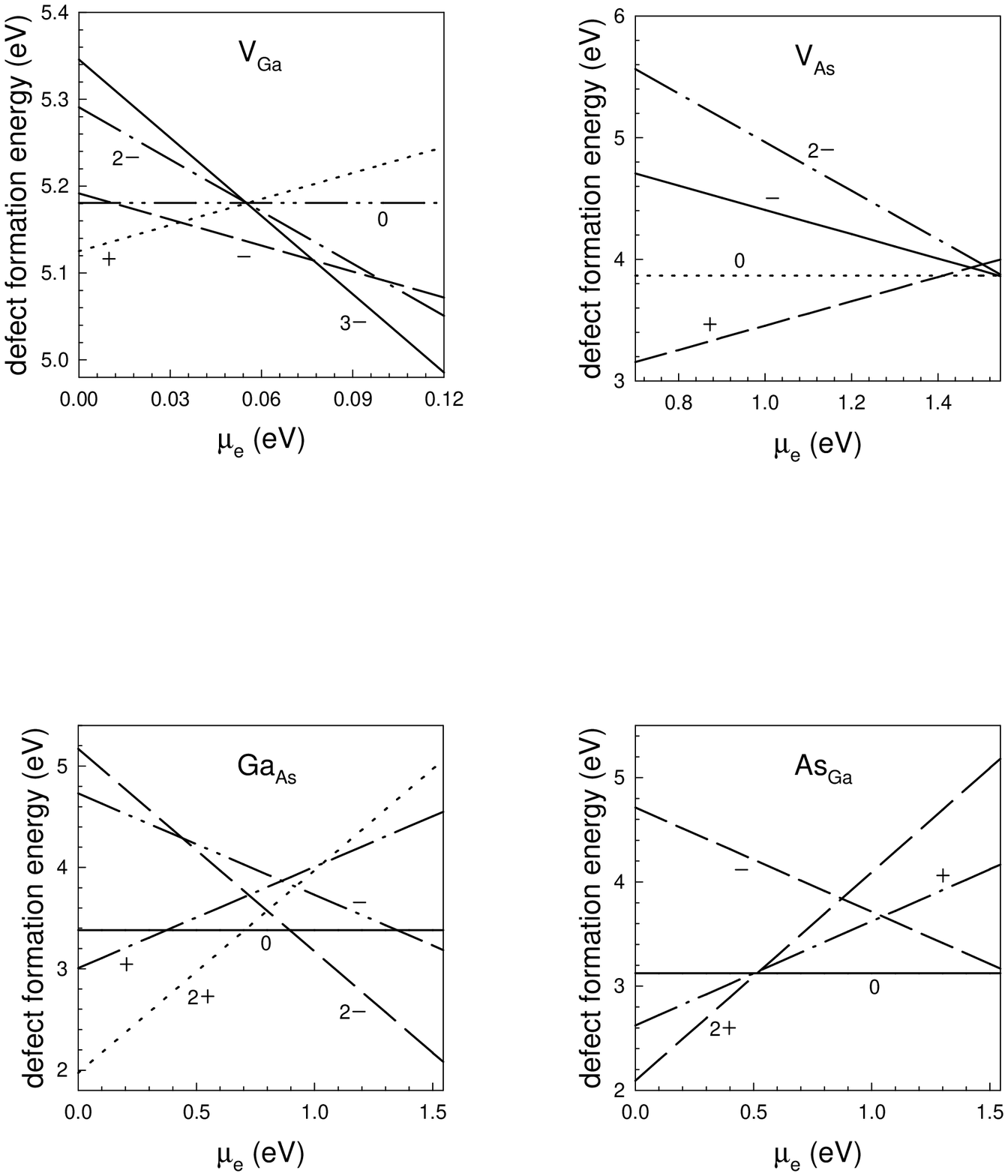}
\vspace{-4cm}
\begin{figure}
\caption{
Formation energies vs electron chemical potential for various defects in GaAs
in different states of charge, as indicated. Here, $\Delta\mu_{\rm Ga} =
-0.85 $ eV. (From Ref.\ \protect\cite{seo95}; reproduced by kind
permission).}
\label{gaasdfe}
\end{figure}

\newpage
\epsfxsize=12cm
\epsfbox{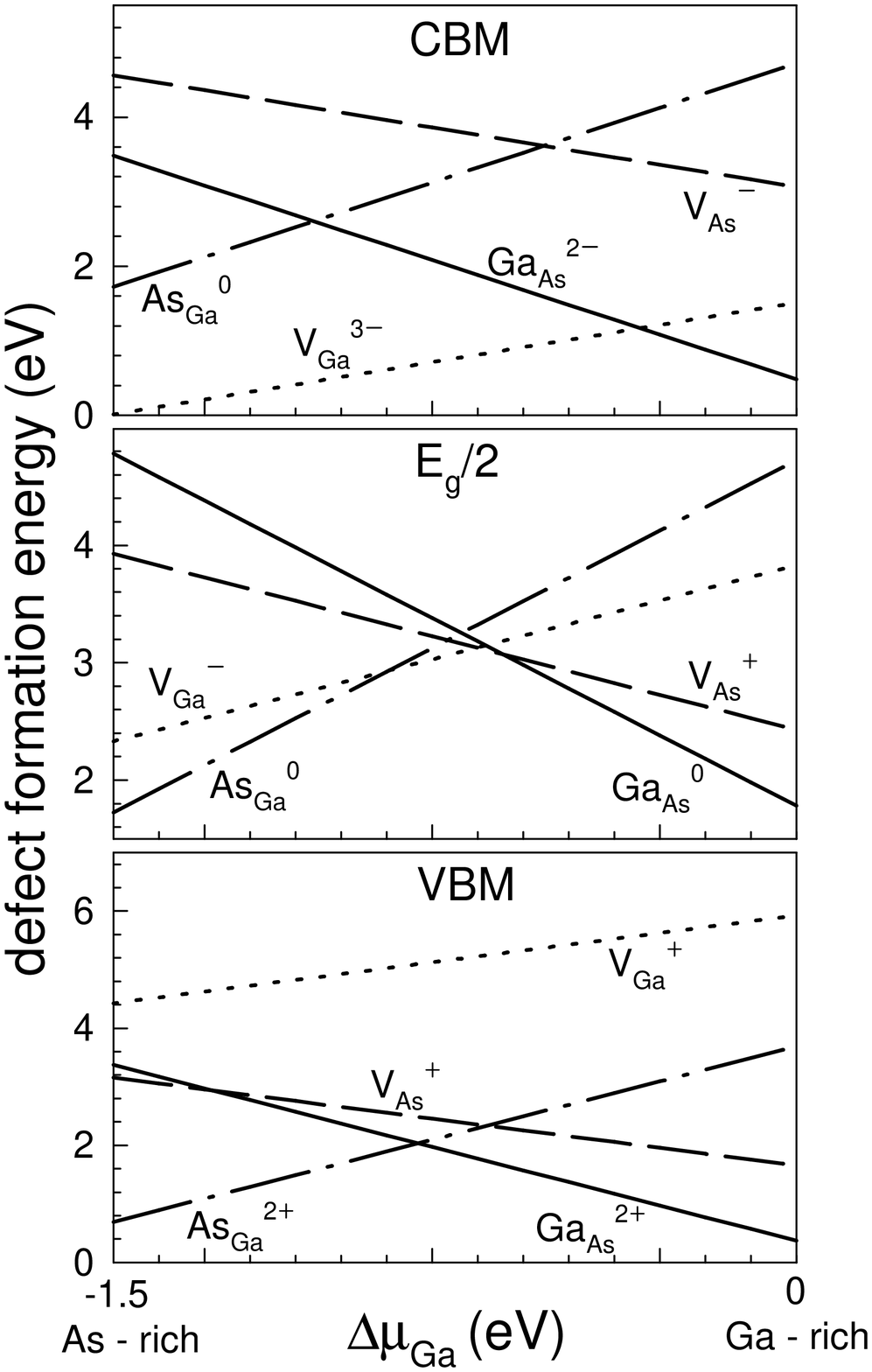}
\vspace{1cm}
\begin{figure}
\caption{
Defect formation energies in GaAs vs $\Delta\mu_{\rm Ga}$ for three different
values of the electron chemical potential: valence-band maximum (VBM), midgap
($E_g/2$), and conduction band minimum (CBM). The lower and upper limits of
the $\Delta\mu_{\rm Ga}-$range correspond to As and Ga-rich regions,
respectively. (From Ref.\ \protect\cite{seo95}; reproduced by kind
permission).}
\label{df_mga}
\end{figure}

\newpage
\epsfxsize=18cm
\epsfbox{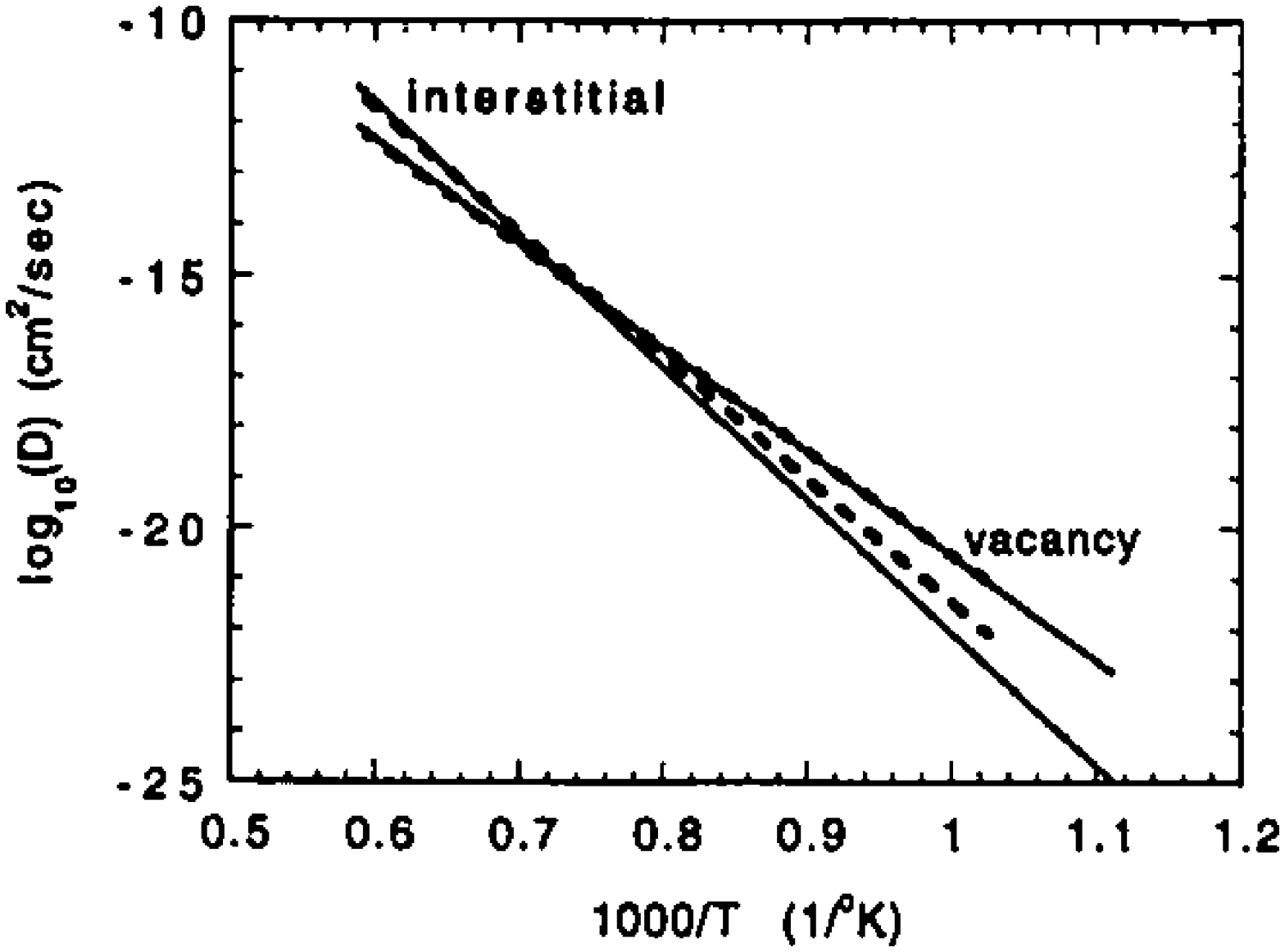}
\vspace{1cm}
\begin{figure}
\caption{
Arrhenius plot of the diffusion constant for vacancies and interstitials in
crystalline silicon. (From Ref.\ \protect\cite{tan97}; reproduced by kind
permission).}
\label{ivdiff}
\end{figure}

\newpage
\epsfxsize=18cm
\epsfbox{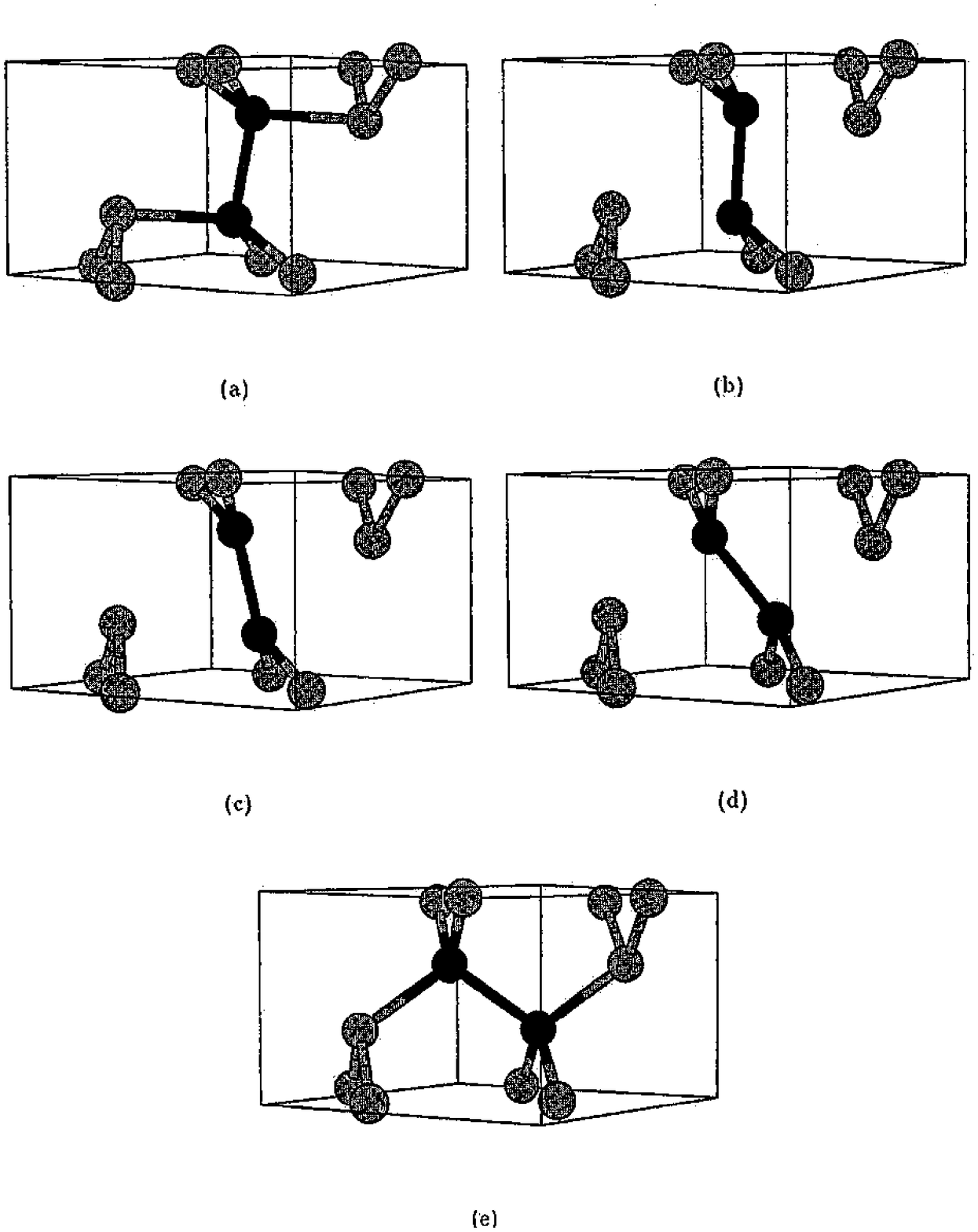}
\vspace{1cm}
\begin{figure}
\caption{
Annihilation of an interstitial-vacancy complex in crystalline silicon: (a)
initial quenched-in configuration; (b) two long bonds are broken; (c)
saddle-point configuration; (d) two new bonds are about to form; (e) final,
defect-free configuration. (From Ref.\ \protect\cite{tan97}; reproduced by
kind permission).}
\label{ivcomp}
\end{figure}

\newpage
\epsfxsize=15cm
\epsfbox{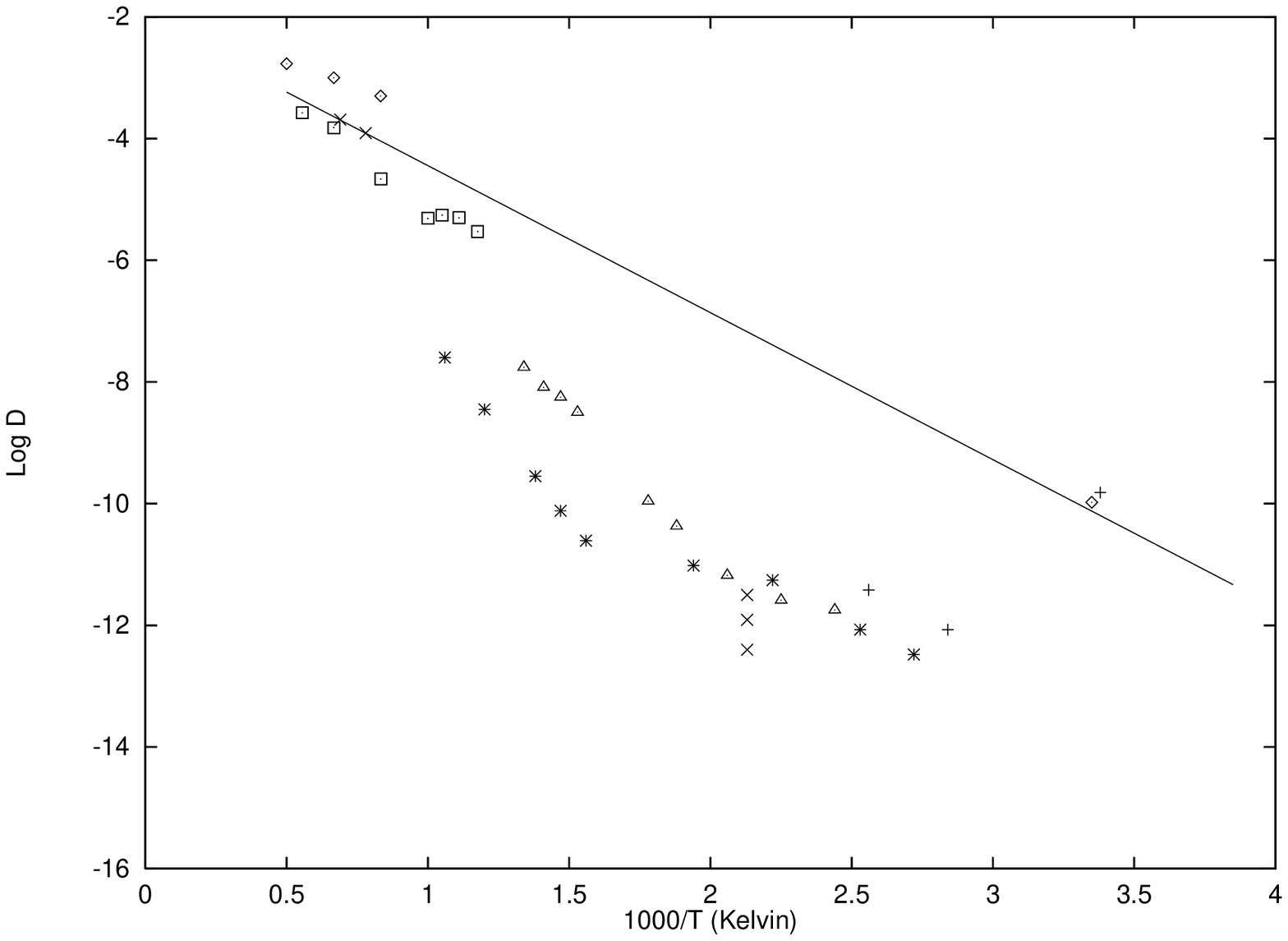}
\vspace{-2cm}
\begin{figure}
\caption{
Arrhenius plot of the diffusion constant of H in {\em c}-Si. The open squares
are the TBMD results of Panzarini and Colombo\protect\cite{pan94} and the
open diamonds are the {\em ab initio} results of Buda {\em et
al.}\protect\cite{bud89}; also given are the corresponding experimental data
(cf.\ \protect\cite{pan94} for references; reproduced by kind permission).}
\label{hsidiff}
\end{figure}

\newpage
\epsfxsize=18cm
\epsfbox{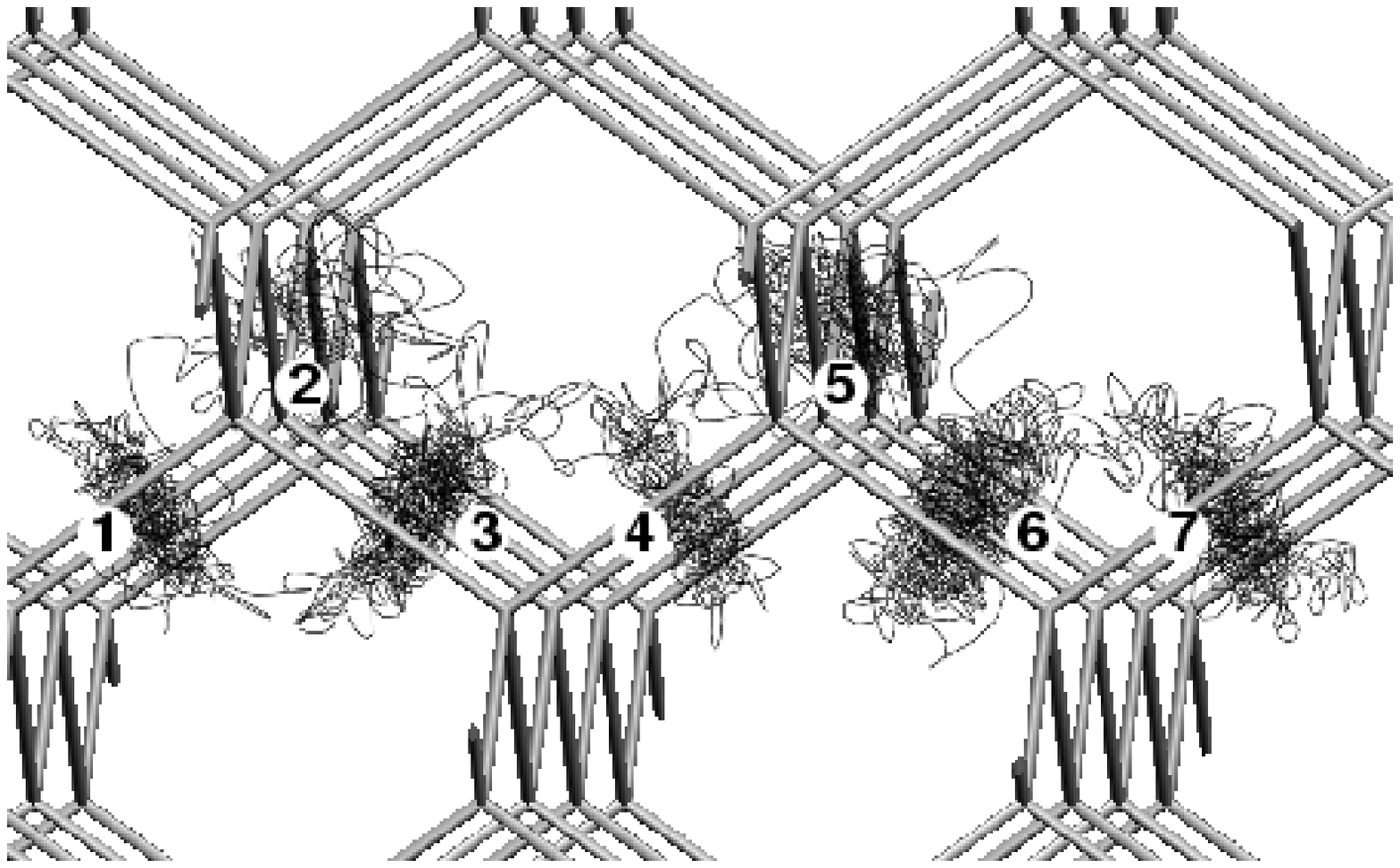}
\begin{figure}
\caption{
Diffusion path of H in {\em c}-Si. (From Ref.\ \protect\cite{pan94};
reproduced by kind permission).}
\label{hsipath}
\end{figure}

\newpage
\begin{figure}
\epsfxsize=18cm
\epsfbox{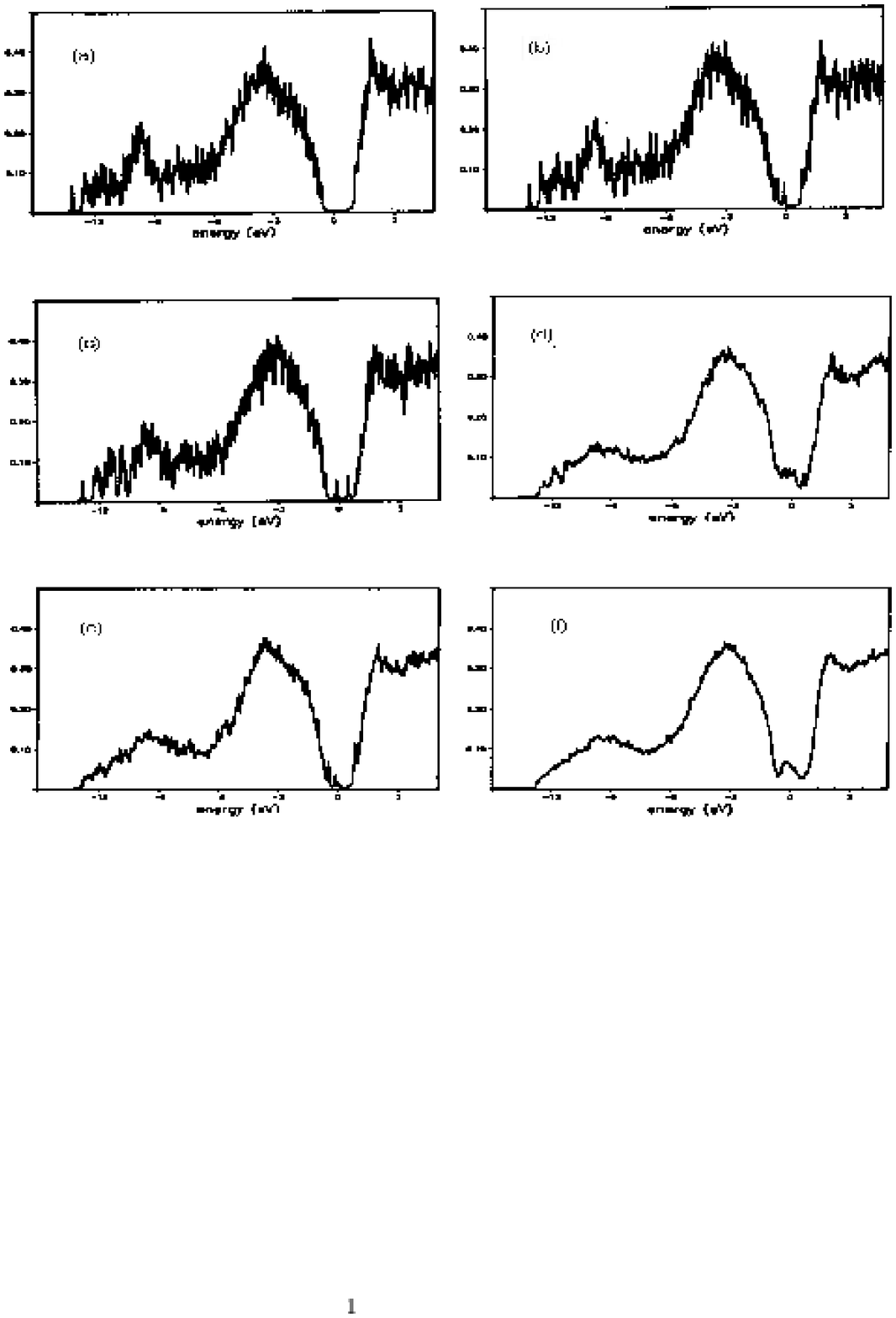}
\vspace{-5cm}
\caption{
TB electronic density of states (spin/eV/atom) for various {\em a-}Si models
obtained using empirical potentials. (a) 216-atom Wooten-Weiner-Weaire (WWW)
model (no coordination defects); (b) 216-atom WWW model relaxed with a
Stillinger-Weber potential (no coordination defects); (c) 216-atom WWW/MD
model (two three-fold defects); (d) 1728-atom model (4\% five-fold and 2\%
three-fold defects); (e) 1728-atom WWW/MD model (no coordination defects);
(f) 13824-atom WWW/MD model (4.5\% five-fold and 2\% three-fold defects).
(From Ref.\ \protect\cite{holender92}; reproduced by kind permission).}
\label{fig:holender}
\end{figure}

\newpage
\begin{figure}
\epsfxsize=18cm
\epsfbox{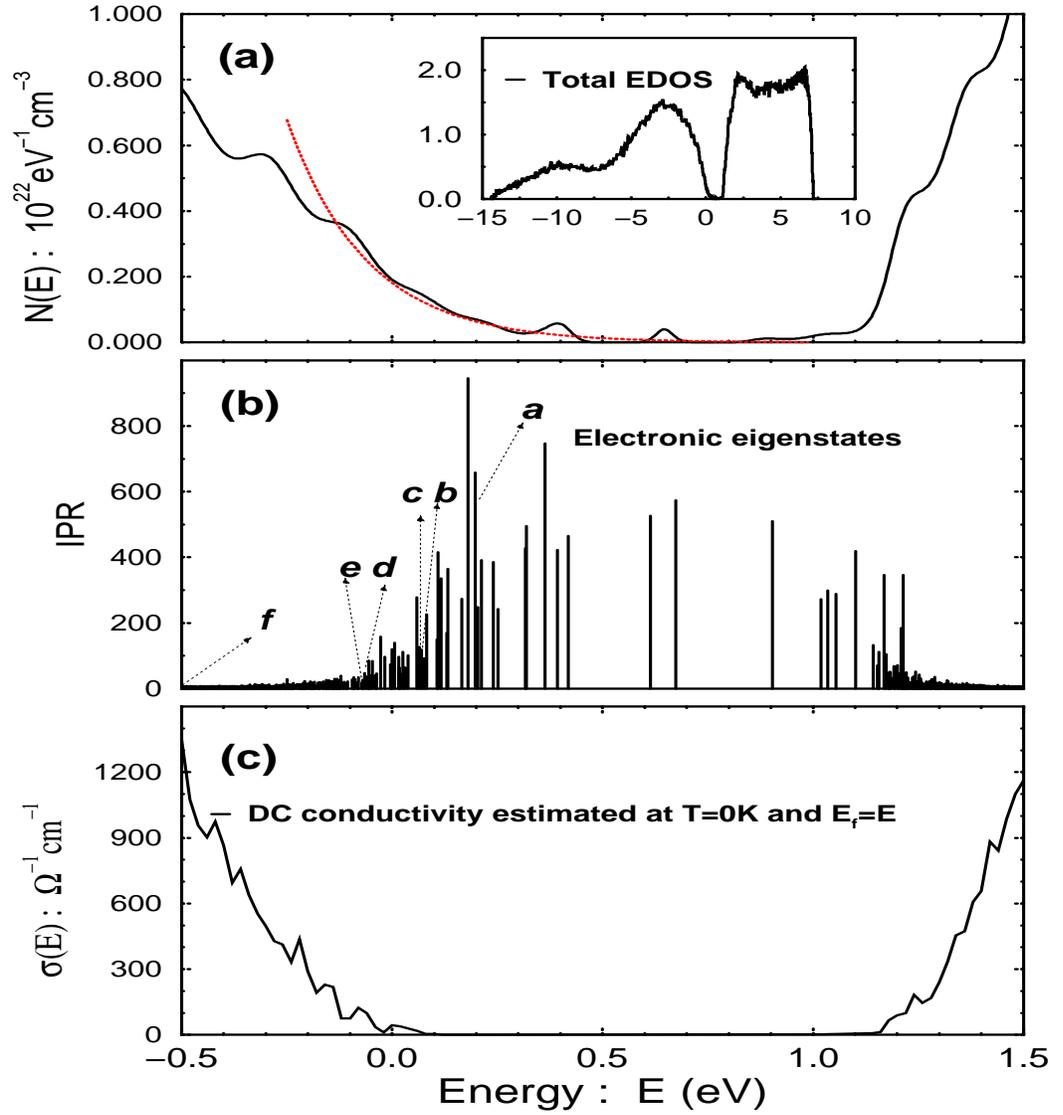}
\vspace{-5cm}
\caption{
Electronic states in the band-gap region of {\em a-}Si: (a) Electronic
density of states; inset: total density of states. (b) Inverse participation
ratio (a measure of localisation) vs energy for states close to the gap. (c)
DC conductivity as a function of doping (computed in the Kubo formalism).
(From Ref.\ \protect\cite{dong98}, reproduced by kind permission.)}
\label{fig:bandtail}
\end{figure}

\newpage
\begin{figure}
\epsfxsize=18cm
\epsfbox{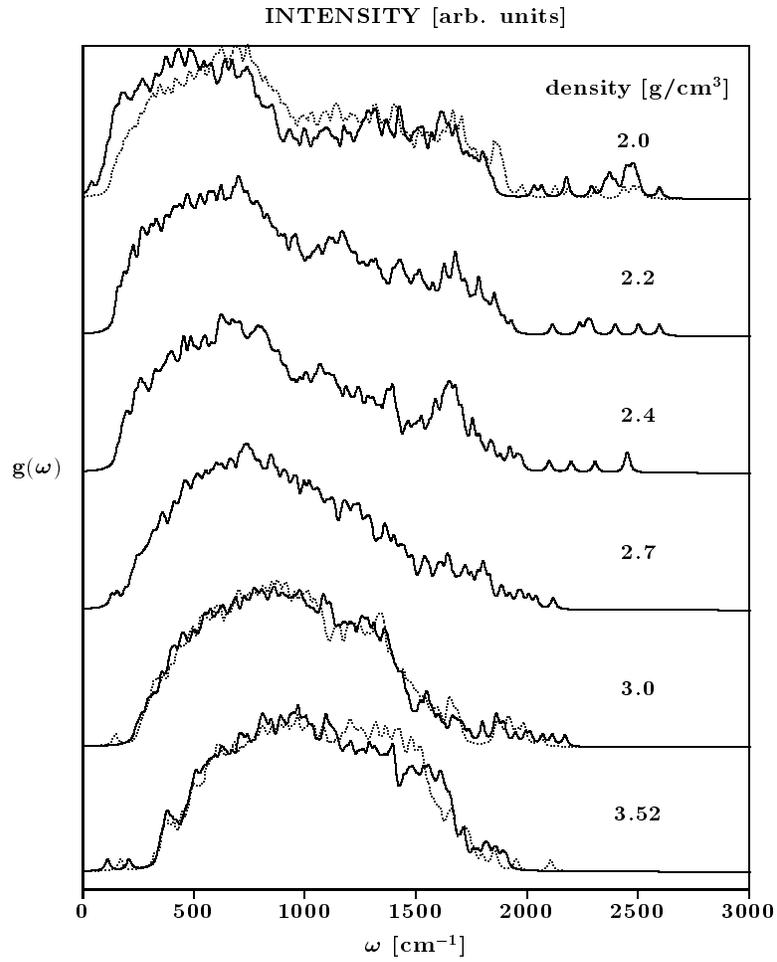}
\vspace{-5cm}
\caption{
Total vibrational density of states (VDOS) for 128-atom cells of amorphous
carbon prepared at different densities. The solid line is for quenched
samples and the dotted line are for samples further annealed at low
temperature. (From Ref.\ \protect\cite{kohler95}; reproduced by kind
permission).}
\label{fig:vdos-ac}
\end{figure}

\newpage
\begin{figure}
\vspace{-2cm}
\epsfxsize=12cm
\epsfbox{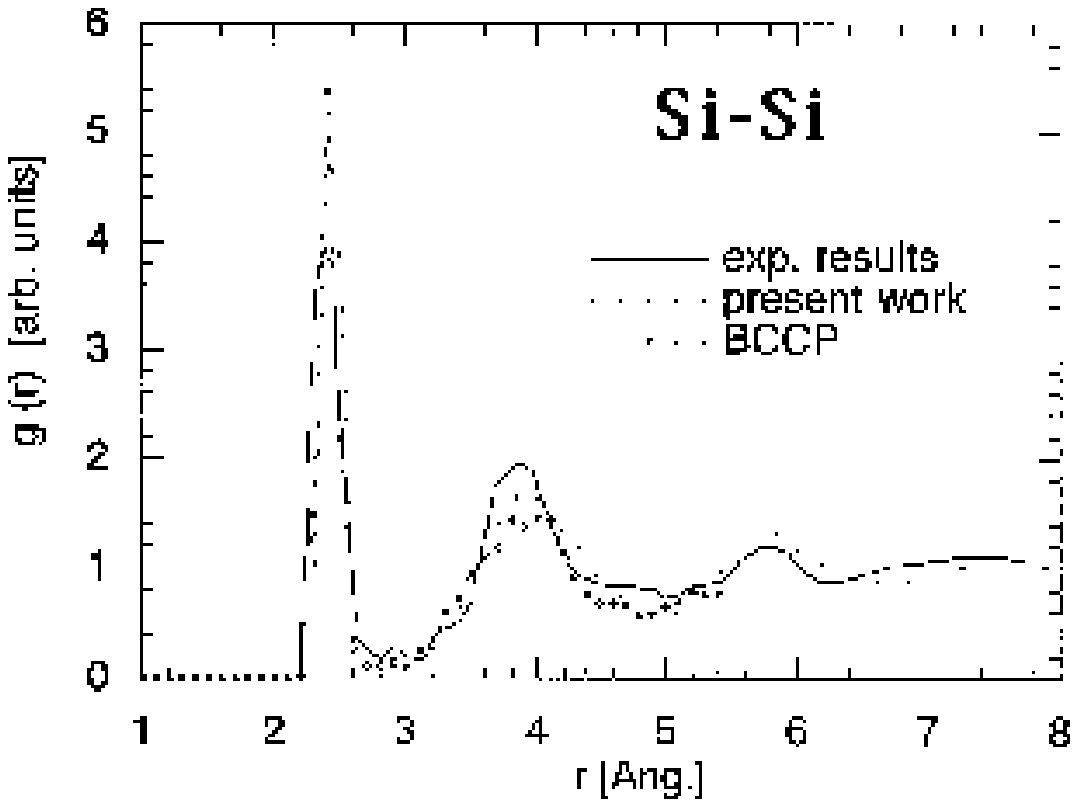}
\vspace{-10cm}
\epsfxsize=12cm
\epsfbox{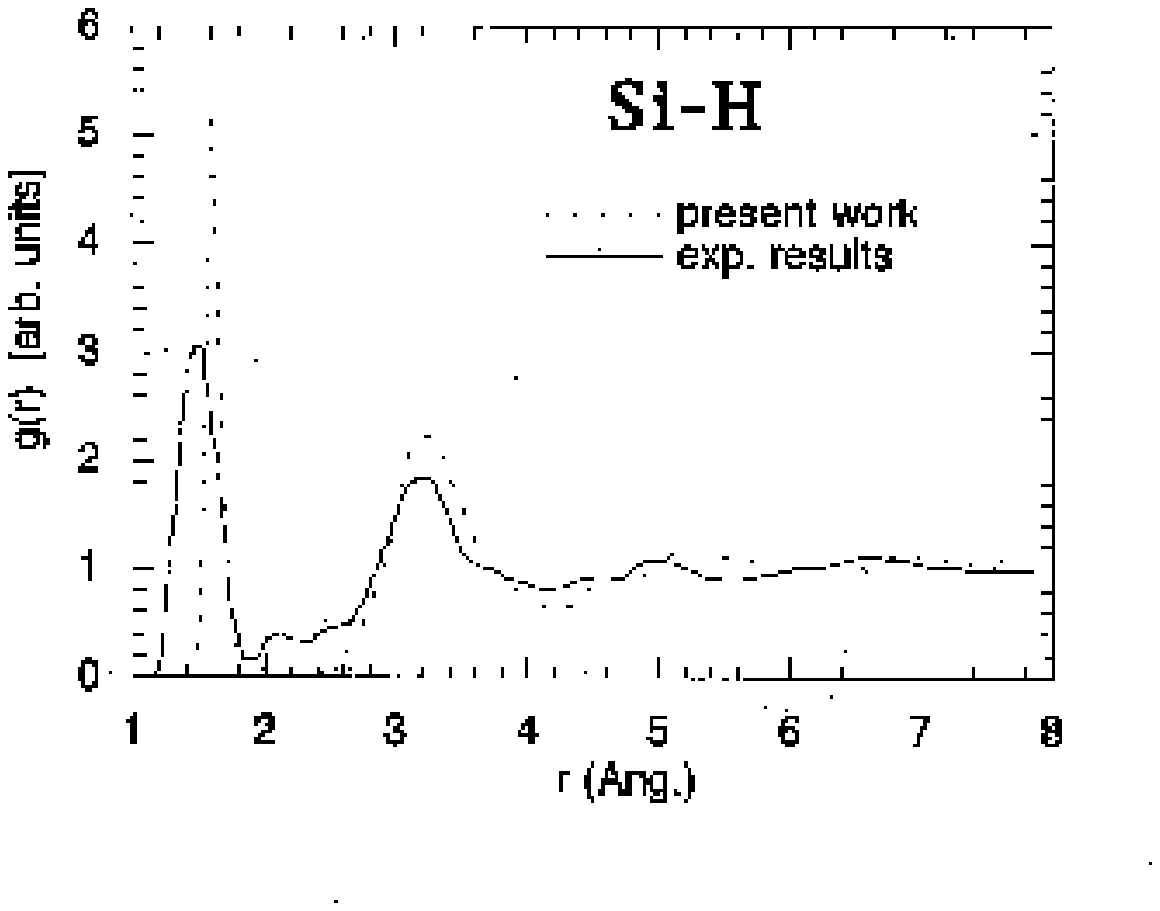}
\vspace{-5cm}
\caption{
Partial radial distribution functions for hydrogenated amorphous silicon.
Full line: experimental measurements of Menelle \protect\cite{menelle87};
dashed line: 216 Si + 26 H model of Tuttle and Adams\protect\cite{tuttle96};
dotted line: {\it ab initio} simulation of the (61 Si + 11 H)-atom cell of
Buda {\it et al.} \protect\cite{buda91}. (From Ref.\ \protect\cite{tuttle96};
reproduced by kind permission).}
\label{fig:asihrdf}
\end{figure}

\newpage
\begin{figure}
\epsfxsize=18cm
\epsfbox{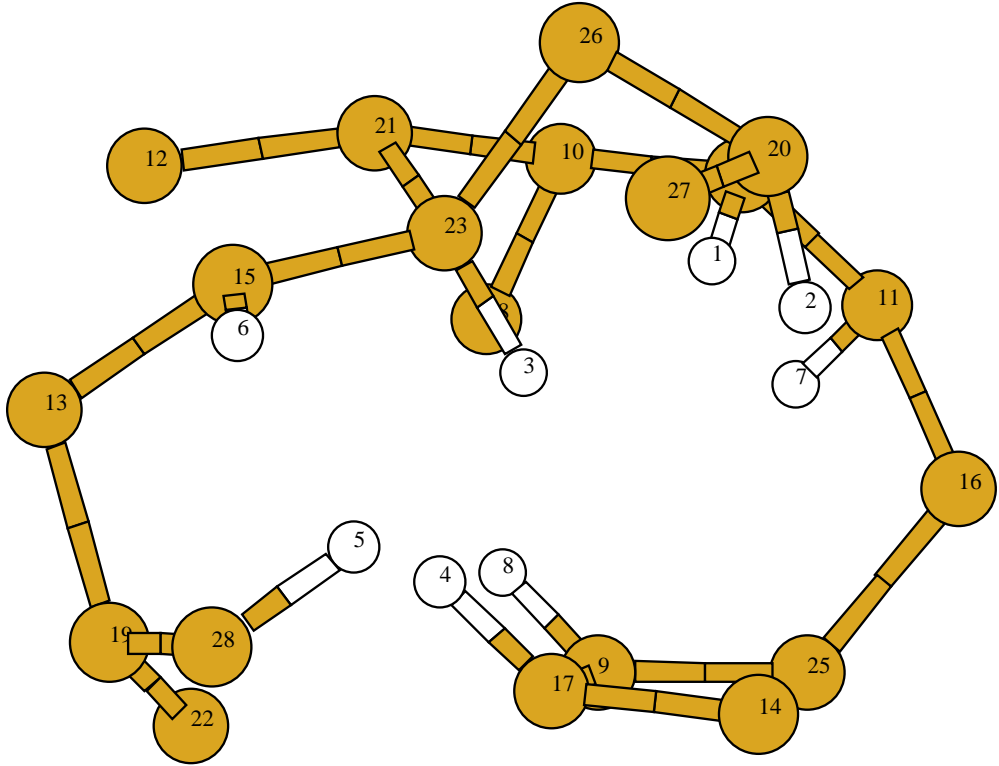}
\caption{
Ellipsoidal cavity in a 242-atom {\it a}-Si:H model containing 26 H, prepared
with MD from a liquid configuration. Eight H atoms bonded to monohydride Si
form the surface of this cavity of dimensions $10\times5\times5$ \AA$^3$.
(From Ref.\ \protect\cite{tuttle96}; reproduced by kind permission).}
\label{fig:cavity}
\end{figure}

\newpage
\vspace*{-8cm}
\begin{figure}
\epsfxsize=18cm
\epsfbox{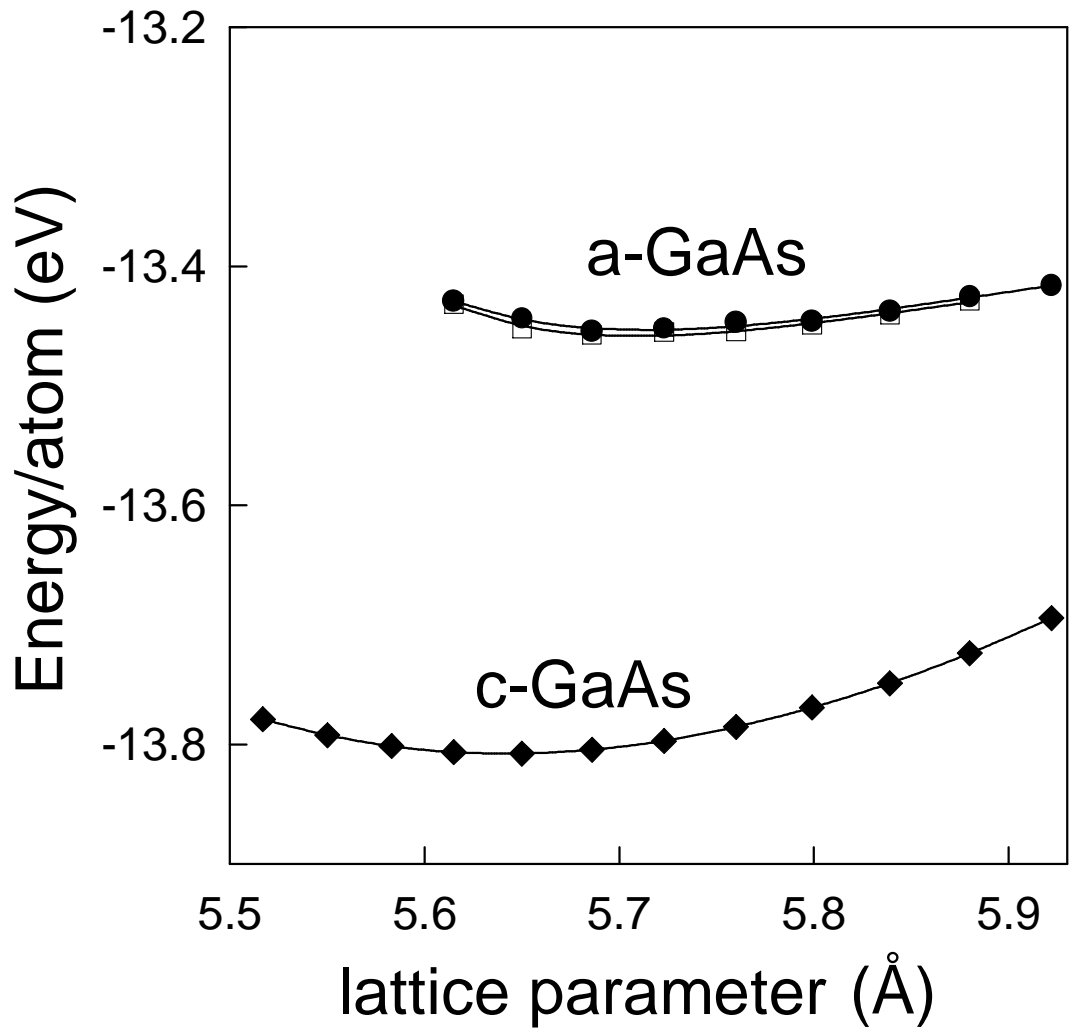}
\caption{
Total energy vs. lattice constant at 0 K for crystalline GaAs and the 64-atom
amorphous samples of Seong and Lewis \protect\cite{seong96}. (Reproduced by
kind permission).}
\label{fig:density}
\end{figure}

\newpage
\begin{figure}
\vspace*{-3cm}
\epsfxsize=18cm
\epsfbox{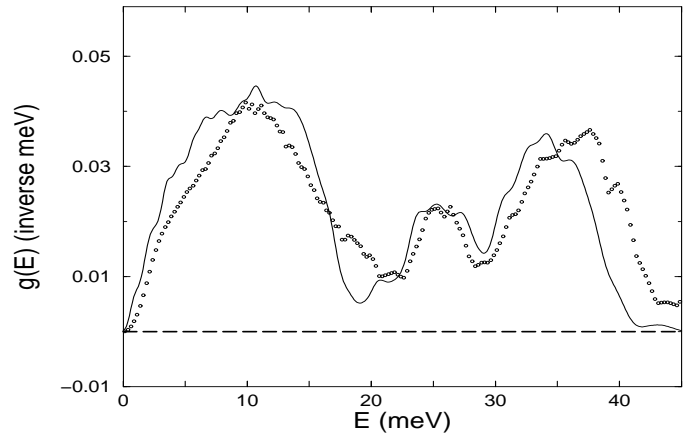}
\vspace{-5cm}
\caption{
Vibrational density of states for amorphous GeSe$_2$. The solid curve is
obtained from the 216-atom cell of Cobb {\it et al.} \protect\cite{cobb96}
and the dots are from experiment \protect\cite{cappelletti95}. (From Ref.\
\protect\cite{cobb96}; reproduced by kind permission).}
\label{fig:vdos}
\end{figure}

\newpage
\begin{figure}
\epsfxsize=18cm
\epsfbox{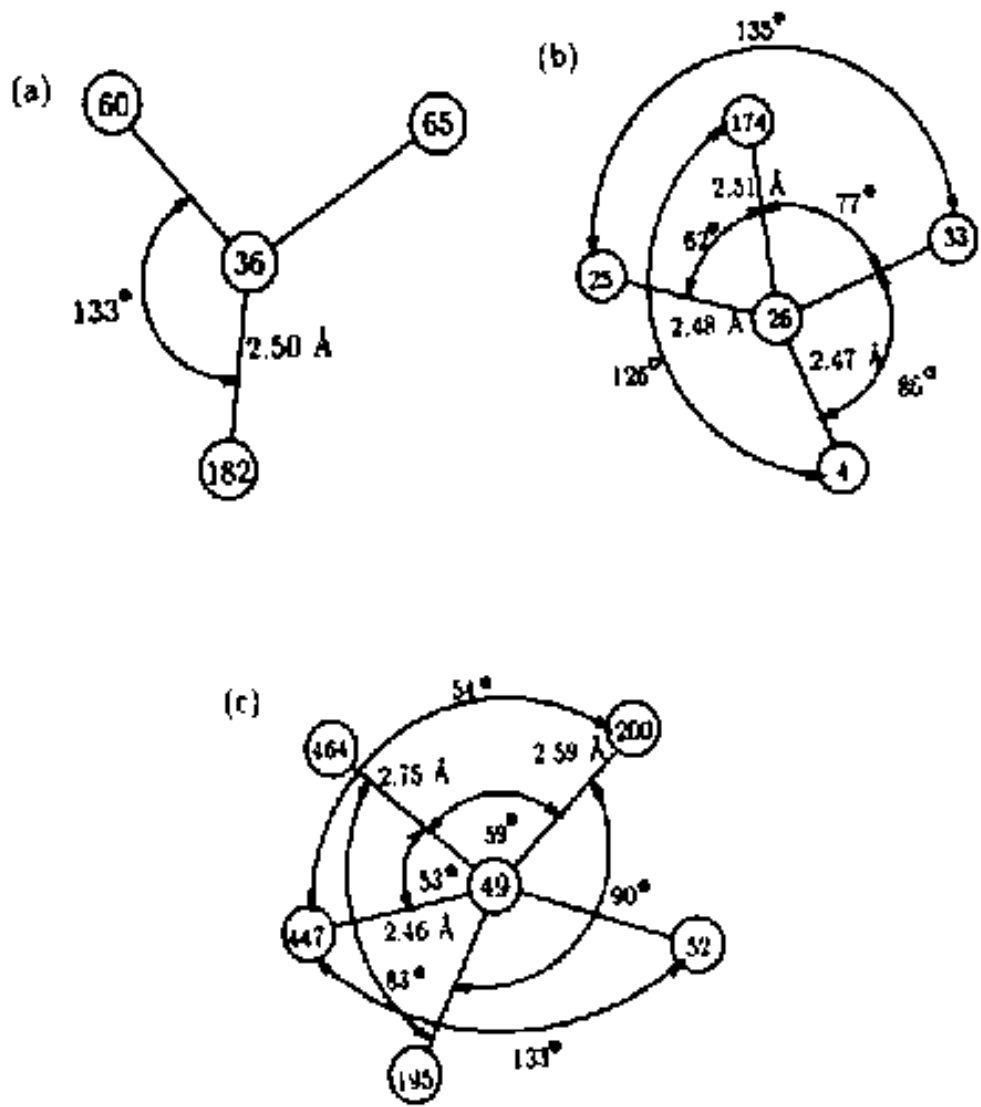}
\vspace{-5cm}
\caption{
Sketches of electronic defects on the {\it a}-Si:H surface . (From Ref.
\protect\cite{kilian93}; reproduced by kind permission).}
\label{fig:sketches}
\end{figure}

\newpage
\begin{figure}
\epsfxsize=18cm
\epsfbox{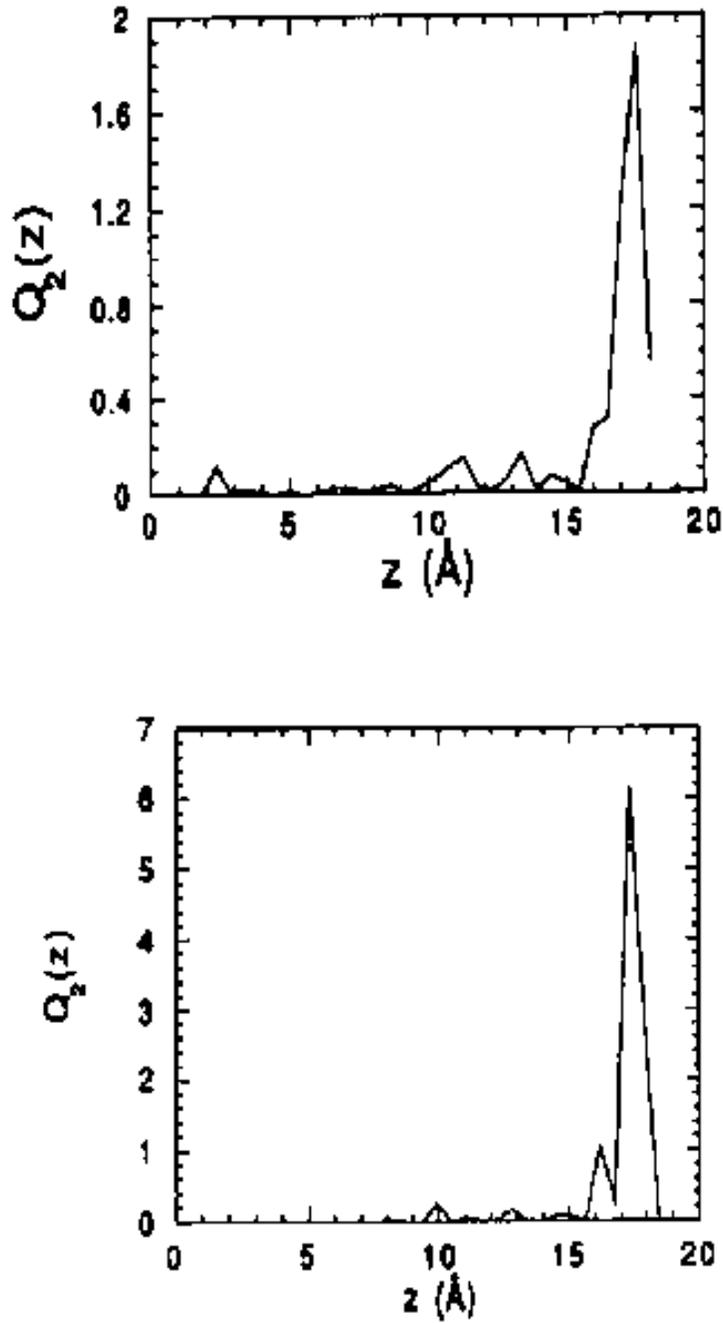}
\vspace{-2cm}
\caption{ Spatial charge localisation as a function of distance from the
bottom layer for defects (b) of Fig.\ \protect\ref{fig:sketches}, before
(top) and after (bottom) hydrogenation of the surface. It can be seen that
the electronic defects do not disappear upon hydrogenation but merely
localise somewhat closer to the surface. (From Ref.\ \protect\cite{kilian93};
reproduced by kind permission).}
\label{fig:Q2H}
\end{figure}

\end{document}